\documentclass[%
 reprint,
 superscriptaddress,
showpacs,
showkeys,
nofootinbib,
 amsmath,amssymb,
 aps,
 prd, twocolumn,
floatfix,
]{revtex4-1}

\usepackage{graphicx}
\usepackage{dcolumn}
\usepackage{bm}

\usepackage{mathtools}
\usepackage{siunitx}
\usepackage{amssymb,multirow,tabularx}
\usepackage{amsmath}
\usepackage{comment}
\usepackage{BibDef}
\usepackage{xcolor}
\usepackage[normalem]{ulem}
\usepackage{hyperref}
\hypersetup{
	colorlinks=true,        
	linkcolor=blue,         
	citecolor=blue,         
}

\newcommand{\msun}{~\rm M_{\large \odot}}
\newcommand{\ud}{~{\rm d}}

\begin{document}

\preprint{APS/123-QED}

\title{Gravitational wave background from extreme mass ratio inspirals}


\author{Matteo Bonetti}
\email[E-mail: ]{matteo.bonetti@unimib.it}
\affiliation{%
 Dipartimento di Fisica ``G. Occhialini'', Universit\`a degli Studi di Milano-Bicocca, Piazza della Scienza 3, 20126 Milano, Italy
}%
\affiliation{%
 INFN, Sezione di Milano-Bicocca, Piazza della Scienza 3, 20126 Milano, Italy
}%

\author{Alberto Sesana}
\affiliation{%
 Dipartimento di Fisica ``G. Occhialini'', Universit\`a degli Studi di Milano-Bicocca, Piazza della Scienza 3, 20126 Milano, Italy
}%
\affiliation{%
 INFN, Sezione di Milano-Bicocca, Piazza della Scienza 3, 20126 Milano, Italy
}%

\received[Received 6 August 2020]{}
\accepted[accepted 2 November 2020]{}
\begin{abstract}
    Extreme mass ratio inspirals (EMRIs), i.e. binary systems comprised by a compact stellar-mass object orbiting a massive black hole, are expected to be among the primary gravitational wave (GW) sources for the forthcoming LISA mission. The astrophysical processes leading to the formation of such systems still remain poorly understood, resulting into large uncertainties in the predicted cosmic rate of these sources, spanning at least three orders of magnitude. As LISA can individually resolve mostly EMRIs up to $z\gtrsim1$, the ensemble of signals below its detection threshold will add up incoherently forming an unresolved confusion noise, which can be formally described as a stochastic background. We perform an extensive study of this background by considering a collection of astrophysically motivated EMRI formation scenarios, spanning current uncertainties. We find that, for most astrophysical models, this signal is easily detectable by LISA, with signal to noise ratios of several hundreds. In fiducial EMRI models -- predicting hundreds of EMRI detections during mission operations -- the background level is comparable to the LISA noise, affecting the performance of the instrument around 3 mHz. In extreme cases, this background can even ``erase'' the whole LISA sensitivity bucket in the 2-10 mHz frequency range. This points to the need of a better understanding of EMRIs' astrophysics for a full assessment of the LISA mission potential.
\end{abstract}


\maketitle

\section{Introduction}

Galactic nuclei are among the densest structures in the Cosmos. They generally host a massive black hole (MBH) at the very center \citep{2013ARA&A..51..511K} and feature a rich content of cold gas, stars and compact objects. Stellar densities in the central parsec can reach $10^6\msun$ pc$^{-3}$, making them an ideal environment for a wide variety of spectacular phenomena driven by extreme dynamics such as stellar tidal disruptions \citep{1988Natur.333..523R}, hypervelocity stars ejection \citep{1988Natur.331..687H} and relativistic captures of compact objects \citep{Amaro-Seoane2018}. Because of dynamical relaxation, close encounters and other dynamical processes, compact objects (COs, generally stellar mass black holes or neutron stars) can be deflected on very low angular momentum orbits, being dynamically captured by the central MBH. 
If specific conditions are met, the CO decouples from the dense stellar environment and together with the MBH evolves as a relativistic binary \citep{Amaro-Seoane2007}. The orbital energy of the binary is gradually released via gravitational wave (GW) emission, causing the slow adiabatic inspiral of the CO onto the MBH. Because of the large disparity in mass of the two object (typically $1-50\msun$ for the CO vs $10^4-10^9\msun$ for the MBH), those events go under the name of extreme mass ratio inspirals \citep[EMRIs,][hereinafter BC04a]{2004PhRvD..69h2005B} and are anticipated to be among the primary GW sources for the  planned space-borne Laser Interferometer Space Antenna \citep[LISA,][]{2017arXiv170200786A}. 

LISA will observe EMRIs at typical GW frequencies lying in the milli-Hz range, which selects 
systems involving MBHs with mass in the interval $10^5$--$10^7 \rm M_\odot$. Due to their slow evolution, these sources typically remain in band for the whole duration of the mission (currently planned to be 4 years), completing up to $\sim 10^5$ orbital cycles before eventually plunging onto the central MBH. The resulting gravitational waveform is very sensitive to the parameters of the EMRI (e.g. masses of the two objects, spin of the MBH, orbital inclination and eccentricity) as well as putative external disturbances from e.g. stellar encounters, or the presence of a dense gaseous disc \citep{2011PhRvD..84b4032K} or a central concentration of dark matter \citep{2020PhRvD.102h3006K}. Therefore EMRIs are extraordinary tools for mapping the space-time around MBHs, promising unprecedented tests of General Relativity (GR) as well as precious insights in the dynamics of dense nuclei \citep{2007PhRvD..75d2003B,Gair2010,Merritt2011,Gair2013,Barausse2014}.

Forecasting EMRI detection prospects for LISA is no easy task. As already mentioned, from a theoretical perspective, EMRIs are expected to form in the center of galaxies, where COs therein can be scattered and directed towards the MBH as a direct consequence of several two-body encounters catalyzed by the large densities of the nuclear regions (i.e. two-body relaxation). A successful EMRI is usually captured onto an highly eccentric orbit, with the subsequent evolution primarily dominated by GW emission \citep{Hils1995,Sigurdsson1997}. 
Several variants of the above process have been proposed so far, either considering modification of the picture, by adding further physical effects such as resonant relaxation and BH-BH scattering events, or invoking different formation processes, including migration of COs in AGN discs, capture by separation of stellar binaries, supernova kicks and more \citep{Levin2003,Miller2005,Hopman2005,Hopman2006,Amaro-Seoane2007,Gair2010,Merritt2011,Mapelli2012,Amaro-Seoane2013,Brem2014,Aharon2016,Bar-Or2016,Babak2017,Chen2018,Amaro-Seoane2018,Bortolas2019}. Even without entering into these complications, in the vanilla capture scenario, the cosmic formation rate of EMRIs depends on a number of poorly known ingredients including the mass function of MBHs below $10^6\msun$ \citep{2019ApJ...883L..18G}, the typical densities of compact objects in galactic nuclei \citep{2011CQGra..28i4017A}, the ratio of successful EMRIs to direct plunges \citep{2015ApJ...814...57M} and many more. Those uncertainties have been investigated by \citep[][hereinafter Babak17]{Babak2017}, who found LISA detection rates spanning three orders of magnitude from just about one to several thousands per year, with fiducial models resulting in a couple of hundred EMRIs per year.

Given the complexity of their waveform \citep[e.g.][]{2006PhRvD..73b4027D,2008PhRvD..77d4013P,2009CQGra..26u3001B,Berry2016,2017PhRvD..96d4005C}, in general, a relatively high signal-to-noise ratio (S/N) of 20 is required for EMRI detection. The extreme mass ratio nature of these systems implies relatively weak GW signals and mainly systems at $z<1$ can reach this S/N threshold. Therefore, it is anticipated that besides the hundreds of observable EMRIs, many thousands more will be present in the LISA data without meeting the detection threshold either because they are too far away or because they are caught too early in their adiabatic inspiral, perhaps hundreds of years far from coalescence. The incoherent piling up of the gravitational radiation emitted by sub-threshold EMRIs could therefore generate an important confusion noise that can be formally described as a stochastic GW background \citep[GWB, Ref.][hereinafter BC04b]{Barack2004}. In the worst case scenario, this signal could even exceed the LISA noise power spectral density (PSD), thus affecting the detectability of other sources. This is, for example, the case with the collective signal from unresolvable Galactic WD binaries, which constitutes the primary limitation of LISA sensitivity to other sources in the frequency range $[0.2,3]\,$mHz \citep{2001A&A...375..890N,2010ApJ...717.1006R}.  

The stochastic GWB from EMRIs has been largely ignored in the literature, and its only systematic computation dates back to the pioneering work of BC04b. Despite the indisputable importance of this seminal work, we are now in the position of improving on their estimates in a number of ways. From the GW signal computation standpoint, BC04b used basic piece-wise approximations for the inclination- and eccentricity-averaged GW signal from unresolved EMRIs. This was combined with early estimates of the EMRI rates, in terms of a redshift independent MBH mass function. To improve upon those assumptions, here we use the EMRI populations of Babak2017, which are constructed by employing a range of physically motivated prescriptions to explore uncertainties due to our current knowledge of MBH evolution and the dynamical processes leading to EMRI formation. From those populations, we extract Montecarlo realizations of EMRIs and compute the GWB from unresolved sources by adding up all individual harmonics of each signal computed exploiting a simplified version of the analytic kludge (AK) waveforms of BC04a, which results in a more accurate estimate of the signal. Finally, the LISA detector underwent a long series of transformations since the early 2000s', resulting in a substantial revision of its noise PSD. Here we specialize our results to the latest LISA sensitivity curve as specified in the ``LISA Science Requirement'' document (referenced as ESA-L3-EST-SCI-RS-001\_LISA\_SciRD
\footnote{See \url{https://atrium.in2p3.fr/nuxeo/nxpath/default/Atrium/sections/Public/LISA/LISA-SciRD-ESA-L3-EST-SC@view_documents?tabIds=\%3A&conversationId=0NXMAIN1} for additional details.}).

The paper is organised as follows: in Section~\ref{sec:method} we describe the developed framework, such as the employed LISA sensitivity curve, the (simplified) Fourier-domain waveform adopted, as well as an operative description of the computation of the GWB. In Section~\ref{sec:results}, we present some estimates of the GWB level when some astrophysically motivated models available in the literature are considered, while in Section~\ref{sec:discussion}, we discuss the possible implications and caveats of the obtained results. Finally, in Section~\ref{sec:conclusions} we draw our conclusions.

\section{Method}
\label{sec:method}

We first describe all the ingredients necessary to the estimation of the GWB from a population of eccentric sources. In particular we have to consider an appropriate waveform model suited for arbitrarily high eccentricity, the sensitivity curve of LISA, that combined with the waveform allows us to evaluate which sources can be individually resolved and therefore that do not contribute to the GWB and, finally, in order to produce sensible estimations of the level and shape of the GWB, we need an astrophysical-based set of catalogues of coalescing EMRIs spanning the wide range of predicted merging rates.
Throughout the paper, unless otherwise stated, we employ the following definitions for the physical quantities needed to characterise the GW system under study:

\footnotetext{Note that, in general, this is different from the mission duration $T_{\rm mission}$ whenever the observation duty cycle ${\cal D}$ is smaller then unity. In fact $T_{\rm obs}={\cal D}\times T_{\rm mission}$. In this paper, we assume continuous LISA observations and therefore $T_{\rm obs}=T_{\rm mission}$.}
\addtocounter{footnote}{-1}
\begin{align*}
   T_{\rm obs} &= \text{total observation time\footnotemark}\nonumber\\
   z &= \text{redshfit}\nonumber\\
   m_1, m_2 &= \text{rest-frame primary and secondary masses}\nonumber\\
   q &= m_2/m_1 \leq 1 = \text{mass ratio}\nonumber\\
   \mathcal{M} &= \dfrac{(m_1 m_2)^{3/5}}{(m_1+m_2)^{1/5}} = \text{rest-frame chirp mass}\nonumber\\
   \mathcal{M}_z &= \mathcal{M}(1+z) = \text{redshifted chirp mass}\nonumber\\
   f &= \text{observed GW frequency}\nonumber\\
   t &= \text{observed time}\nonumber\\
   t_r &= t/(1+z) = \text{rest-frame time}\nonumber\\
   f_r &= f (1+z) = \text{rest-frame GW frequency}\nonumber\\
   f_{\rm orb} &= \text{rest-frame orbital frequency}\nonumber\\
   n &= \text{harmonic number}\nonumber\\
\end{align*}

\begin{align*}
   f_n &= n f_{\rm orb} = \text{$n$-th GW harmonic rest-frame frequency}\nonumber\\
   e_n &= e(f_r/n) = \text{eccentricity at $f_{\rm orb} = f_r/n$}\nonumber\\
   d &= \text{comoving distance}\nonumber\\
   d_L &= d (1+z) = \text{luminosity distance}\nonumber\\
   \dfrac{f_n}{\dot{f}_n} &= \dfrac{\ud t_r}{\ud \ln f_n} = \dfrac{\ud t_r}{\ud \ln f_{\rm orb}} = \text{residence time at $f_n$}
\end{align*}

\subsection{LISA sensitivity}
\label{sec:LISA_sens}

Throughout the paper, we consider a six-link LISA configuration (i.e. one consisting of two independent detectors). We adopt the sky-averaged, LISA sensitivity curve as detailed in the ``LISA Science Requirement Document'' document. 

Besides the instrumental noise we also take into account the effect of a large population of unresolved galactic compact binaries (mostly white-dwarf, WD, binaries).
This population produces a stochastic ``confusion noise'' that effectively degrades the instrumental sensitivity at frequencies below $\sim 1$ mHz. Being dominated by systems located within the Galactic disk, the WD confusion noise is expected to be highly anisotropic and its amplitude in the LISA detector is anticipated to fluctuate due to the satellite constellation motion. This feature makes this signal at least partially subtractable from the LISA error budget when searching for an underlying stochastic background \citep{2014PhRvD..89b2001A}. We do not consider this possibility in the present study, and our results are conservative in this respect. Combining the instrumental and confusion noise contributions, the total (sky-average) LISA sensitivity as a function of frequency $f$ can be express as
\begin{equation}
    S_{\rm noise}(f) = \dfrac{1}{2}\dfrac{20}{3} \left(\dfrac{S_{\rm I}(f)}{(2 \pi f)^4} + S_{\rm II}(f)\right) \times R(f) + S_c(f),
\end{equation}
where 
\begin{align}
    S_{\rm I}(f) &= 5.76 \times 10^{-48}\left(1+\left(\dfrac{f_a}{f}\right)^2\right) \ {\rm s^{-4} Hz^{-1}}\nonumber\\
    S_{\rm II}(f) &= 3.6 \times 10^{-41} \ {\rm Hz^{-1}}\nonumber\\
    R(f) &= 1+\left(\dfrac{f}{f_b}\right)^2 
\end{align}
with $f_a = 0.4$ mHz and $f_b = 25$ mHz. The galactic confusion noise is fitted with the formula (Karnesis \& Babak in preparation)
\begin{equation}
    S_{\mathrm{gal}} = \frac{A}{2} e^{-(f/f_1)^\alpha} f^{-7/3} \left( 1 + \mathrm{tanh}\left( \frac{f_\mathrm{knee} - f}{f_2} \right) \right),
    \label{eq:galfit}
\end{equation}
where $A = 1.28\times 10^{-44}$ is the overall amplitude, $\alpha = 1.63$ is a smoothness parameter, while $f_1$, $f_2$ and $f_\mathrm{knee}$ denote break frequencies that parametrise the model. In particular, those frequencies depend on the observation time and reads
\begin{align}
    \log_{10}\left( f_1/\mathrm{Hz} \right) &= a_1 \log_{10} (T_{\mathrm{obs}}/\mathrm{yr}) + b_1, \nonumber\\
    \log_{10}(f_2/\mathrm{Hz}) &= -3.318 \nonumber\\
    \log_{10}\left( f_{\mathrm{knee}}/\mathrm{Hz}\right) &= a_k \log_{10}(T_{\mathrm{obs}}/\mathrm{yr}) + b_k,
\label{eq:galfittobs}
\end{align}
where $a_1$, $a_k$, $b_1$, and $b_k$ are parameters that depend on the adopted S/N threshold for WD binary detectability. If the threshold is set to ${\rm S/N} = 7$ then:

\begin{align}
    a_1 &= -0.224\nonumber\\
    b_1 &= -2.704\nonumber\\
    a_k &= -0.361\nonumber\\
    b_k &= -2.378.
\end{align}

The dependence of $f_1$ and $f_{\rm knee}$ on $T_{\rm obs}$ implies that the WD confusion noise becomes lower and lower as LISA collects data. This is because LISA's frequency resolution improves as $T_{\rm obs}$ and individual WD S/N grows as $T_{\rm obs}^{1/2}$. Therefore, the longer $T_{\rm obs}$, the larger is the number of individually resolvable WD binaries, leaving behind a lower residual unresolved confusion noise.  

\subsection{Waveforms and S/N calculation}
\label{sec:SNR}

We turn now to the description of the formalism employed to model the inspiral of EMRIs. In the standard astrophysical picture, the successful formation of an EMRI implies the capture of the CO onto an extremely eccentric orbit (circularity $1-e$ around $10^{-5}-10^{-4}$). Despite efficient GW circularization, the system can still retain an high eccentricity (as high as $0.99$) when entering the LISA band. Therefore we need to focus on waveform models that can handle eccentric sources. Moreover,
to compute the GWB, we need to add-up signals from hundreds of thousands of EMRIs, which requires a waveform model that is also fast. To accommodate these requirements, we develop a simplified version of the PN formalism of BC04a, which is well suited for a fast computation of the GW signal from a large population of arbitrarily eccentric EMRIs.

We use the Newtonian fluxes worked out in \citet{Peters1964} to evolve the orbital elements of binary systems, i.e. the orbital frequency (related to the semi-major axis) and eccentricity of a given EMRI. This informs us on which frequency range is spanned by each system during the LISA time window. Specifically, we evolve EMRIs in the GW regime with the orbit-averaged equations \citep{Peters1964}

\begin{align}\label{eq:dfdt_gw}
    \frac{\ud f_{\rm orb}}{\ud t}&=\frac{96 G^{5/3}}{5 c^5} (2\pi)^{8/3} \mathcal{M}^{5/3}\, f_{\rm orb}^{11/3}\, \mathcal{F}(e),\\
    \frac{\ud e}{\ud t}&=-\frac{G^{5/3}}{15 c^5} (2\pi)^{8/3} \mathcal{M}^{5/3}\, f_{\rm orb}^{8/3}\, \mathcal{G}(e),
    \label{eq:dedt_gw}
\end{align}
where $\mathcal{M}$ is the source-frame chirp mass, i.e.
\begin{equation}
    \mathcal{M} = \dfrac{(m_1 m_2)^{3/5}}{(m_1+m_2)^{1/5}},
\end{equation}
while $\mathcal{F}(e)$ and $\mathcal{G}(e)$ are algebraic functions of the eccentricity:
\begin{align}\label{eq:peters}
    \mathcal{F}(e) &= \dfrac{1+73/24 e^2 + 37/96 e^4}{(1-e^2)^{7/2}},
    \\
    \mathcal{G}(e) &= \dfrac{304 e + 121 e^3}{(1-e^2)^{5/2}}.
\end{align}

\begin{figure}
    \centering
    \includegraphics[width=0.48\textwidth]{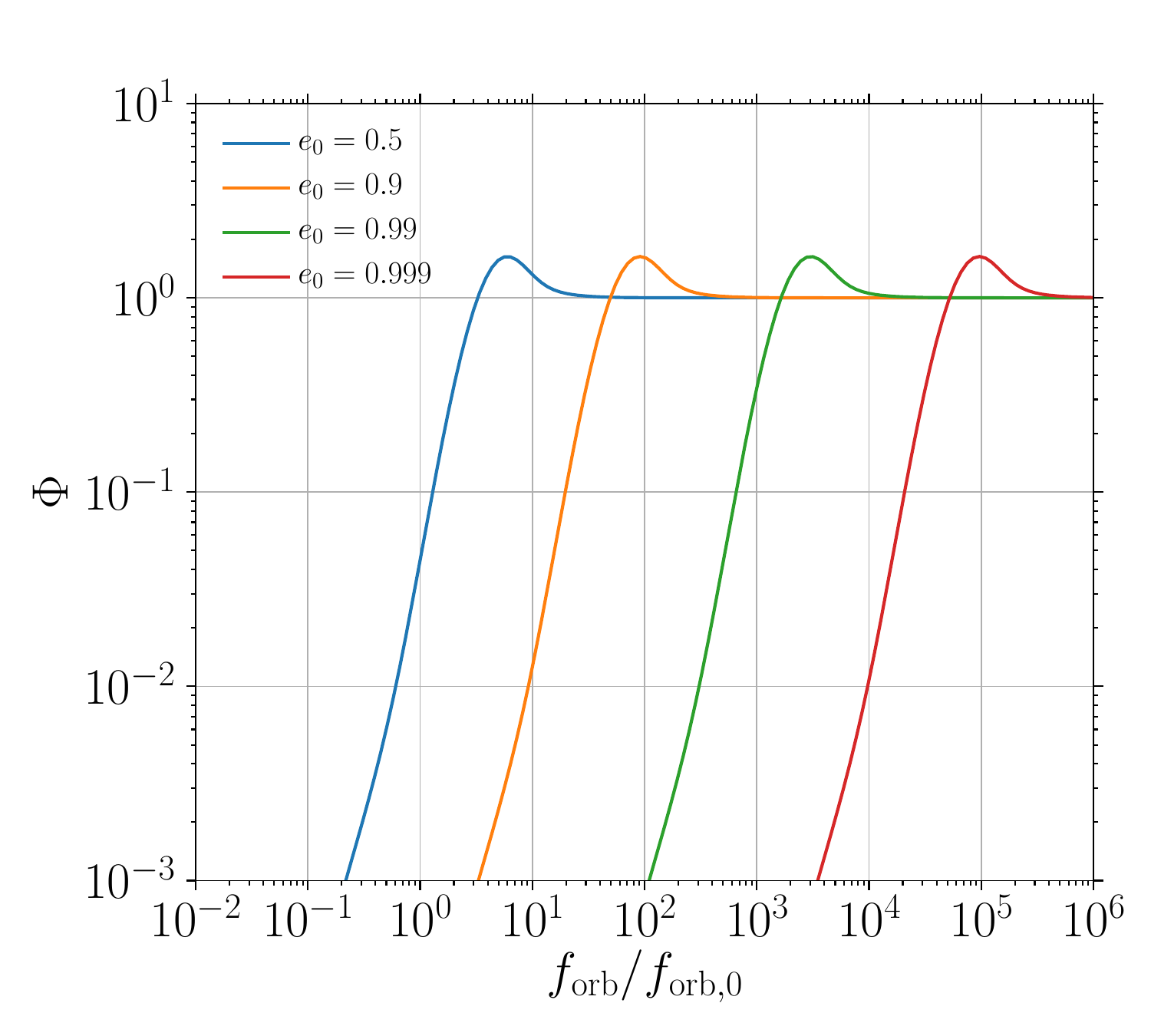}
    \caption{Frequency evolution of the function $\Phi$ for different initial $e_0$ chosen at $f_{\rm orb,0}$.}
    \label{fig:Phi}
\end{figure}

The computation of the S/N implies the knowledge of the emitted waveform. Gravitational radiation emitted by eccentric binaries requires a more complicated treatment compared to the standard circular orbits. While for circular binaries most of the GW power is contained in the dominant quadrupolar mode, whose frequency is twice the orbital frequency, several harmonics are excited with comparable amplitudes in the eccentric case, i.e. the GW spectrum contains several dominant frequencies $f_n = n f_{\rm orb}$, where $n$ is the harmonic number. Therefore the total emitted power in GW is spread on a broad spectrum of frequencies, with the fraction of power per harmonic given by \citep{Peters1963}
\begin{equation}\label{eq:Edot_n}
    \dot{E}_n = \dfrac{32 G^{7/3}}{5 c^5} (2\pi \mathcal{M} f_{\rm orb})^{10/3} g_n(e).
\end{equation}
Here the dimensionless function $g_n(e)$ determines which fraction of the GW power goes into each harmonic and reads
\begin{align}\label{eq:g_n_e}
    g_n(e) &= \frac{n^4}{32} \Bigg[\bigg(J_{n-2}(ne)-2eJ_{n-1}(ne)+\frac{2}{n}J_n(ne)\nonumber\\
    &+2eJ_{n+1}(ne)-J_{n+2}(ne)\bigg)^2 \nonumber\\ &+(1-e^2)\Big(J_{n-2}(ne)-2J_n(ne)+J_{n+2}(ne)\Big)^2 \nonumber\\
    &+ \frac{4}{3n^2} J_n^2(ne)\Bigg],
\end{align}
where $J_n$ represents the $n$-th order Bessel function of the first kind. Note that in the case of a circular binary only the second harmonic contributes, i.e. $g_n(0)=\delta_{2n}$, where $\delta_{mn}$ is the standard Kronecker delta.

We employ the formalism developed in \citet{Finn2000} and BC04a, where the characteristic strain (inclination-polarization averaged) of each harmonic is given by
\begin{equation}\label{eq:hcn}
    h_{c,n} = \dfrac{1}{\pi d} \sqrt{\dfrac{2 G \dot{E}_n}{c^3\dot{f}_n}},\\
\end{equation}
where $\dot{f}_n = n \dot{f}_{\rm orb}$ is the time derivative of the $n$-th harmonic and $d$ is the co-moving distance to the GW source. The total S/N is then computed as 
\begin{equation}\label{eq:SNR}
    ({\rm S/N})^2 = \sum_{n=1}^\infty \int \dfrac{h_{c,n}^2}{f_n S_{\rm noise}(f_n)} \ud \ln f_n,
\end{equation}
where $S_{\rm noise}(f_n)$ is the sky averaged power spectral density of LISA, which according to its definition in Section~\ref{sec:LISA_sens} already accounts for the fact that the LISA constellation is comprised of two independent interferometers.

Despite formally correct, the combination of equation~\eqref{eq:hcn} and equation~\eqref{eq:SNR} turns out to be quite expensive to evaluate, therefore for large EMRI samples the required computational time can be significant. We can note however that each $h_{c,n}$ enters equation~\eqref{eq:SNR} to the square power, thus, by swapping the sum over $n$ and the integral over $f_n$, we can express a ``total characteristic strain'' as the squared sum of the characteristic strains belonging to each harmonic, i.e
\begin{equation}\label{eq:h_tot}
    h_c^2 = \sum_{n=1}^\infty h_{c,n}^2,
\end{equation}
with the only subtlety consisting in evaluating all $h_{c,n}^2$ at the same frequency. In fact, once the observed GW frequency $f$ is fixed, the $n-$th harmonic contributing to the signal at $f$ has to be evaluated when the EMRI is at $f_{\rm orb}=f(1+z)/n$, i.e. at a different evolutionary stage of the EMRI orbit.
Specifically, the first harmonic contributes when the rest-frame orbital frequency is $f_{\rm orb}=f(1+z)$, the second one contributes at $f$ when $f_{\rm orb}=f(1+z)/2$, and so on. 
Thus, further expanding equation~\eqref{eq:hcn} through equation~\eqref{eq:dfdt_gw} and equation~\eqref{eq:Edot_n}, we obtain that at a generic observed GW frequency $f$ the $n$-th characteristic strain reads\footnote{The characteristic strain can be express either in terms of observed ($f,d_L,\mathcal{M}_z$) or rest-frame quantities ($f_r,d,\mathcal{M}$). Here we adopt the first choice.}

\begin{equation}\label{eq:hcn_bis}
    h_{c,n}^2(f) = \dfrac{2 G^{5/3} (2\pi)^{2/3} \mathcal{M}_z^{5/3} f^{-1/3}}{3 c^3 \pi^2 d_L^2} \dfrac{g_n(e_n)}{n^{2/3} \mathcal{F}(e_n)},
\end{equation}
where $d_L = d (1+z)$ is the luminosity distance, $\mathcal{M}_z = \mathcal{M}(1+z)$ is the redshifted chirp mass, while $e_n = e\bigl(f(1+z)/n\bigr)$ is the eccentricity corresponding to a binary rest-frame orbital frequency $f_{\rm orb}=f(1+z)/n$. Finally, summing over all harmonics the total characteristic strain for a single EMRI at a generic $f$ is given by
\begin{align}\label{eq:h_tot2}
    h_c^2(f) &= \dfrac{2 G^{5/3} \pi^{2/3} \mathcal{M}_z^{5/3} f^{-1/3}}{3 c^3 \pi^2 d_L^2} \times \Phi(f), \\
    \Phi(f) &= 2^{2/3} \sum_{n=1}^\infty \dfrac{g_n(e_n)}{n^{2/3} \mathcal{F}(e_n)},
\end{align}
with the only difficulty represented by the evaluation of the sum $\Phi(f)$ over several $n$.

\begin{figure*}
    \centering
    \includegraphics[width=0.48\textwidth]{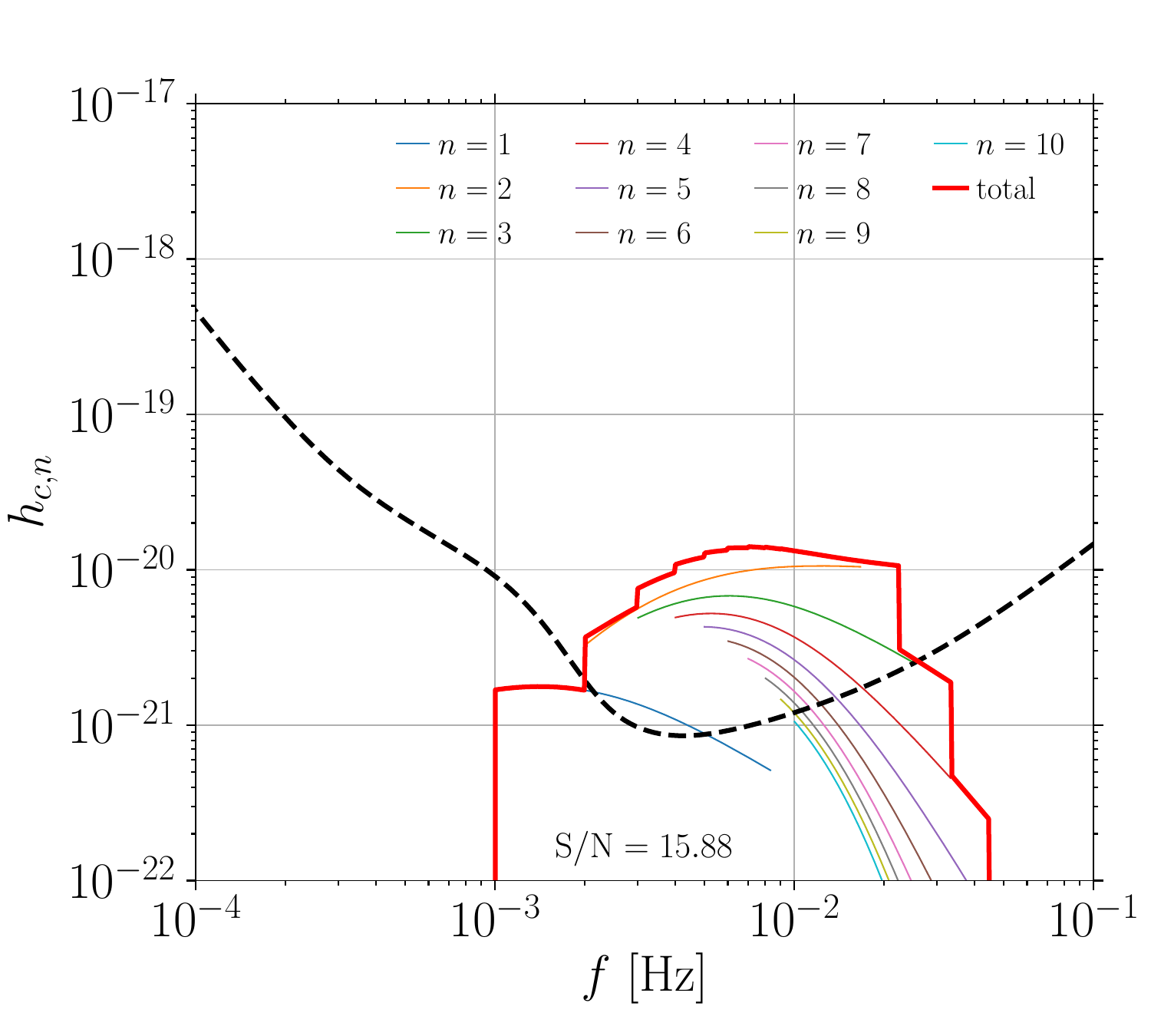}
    \includegraphics[width=0.48\textwidth]{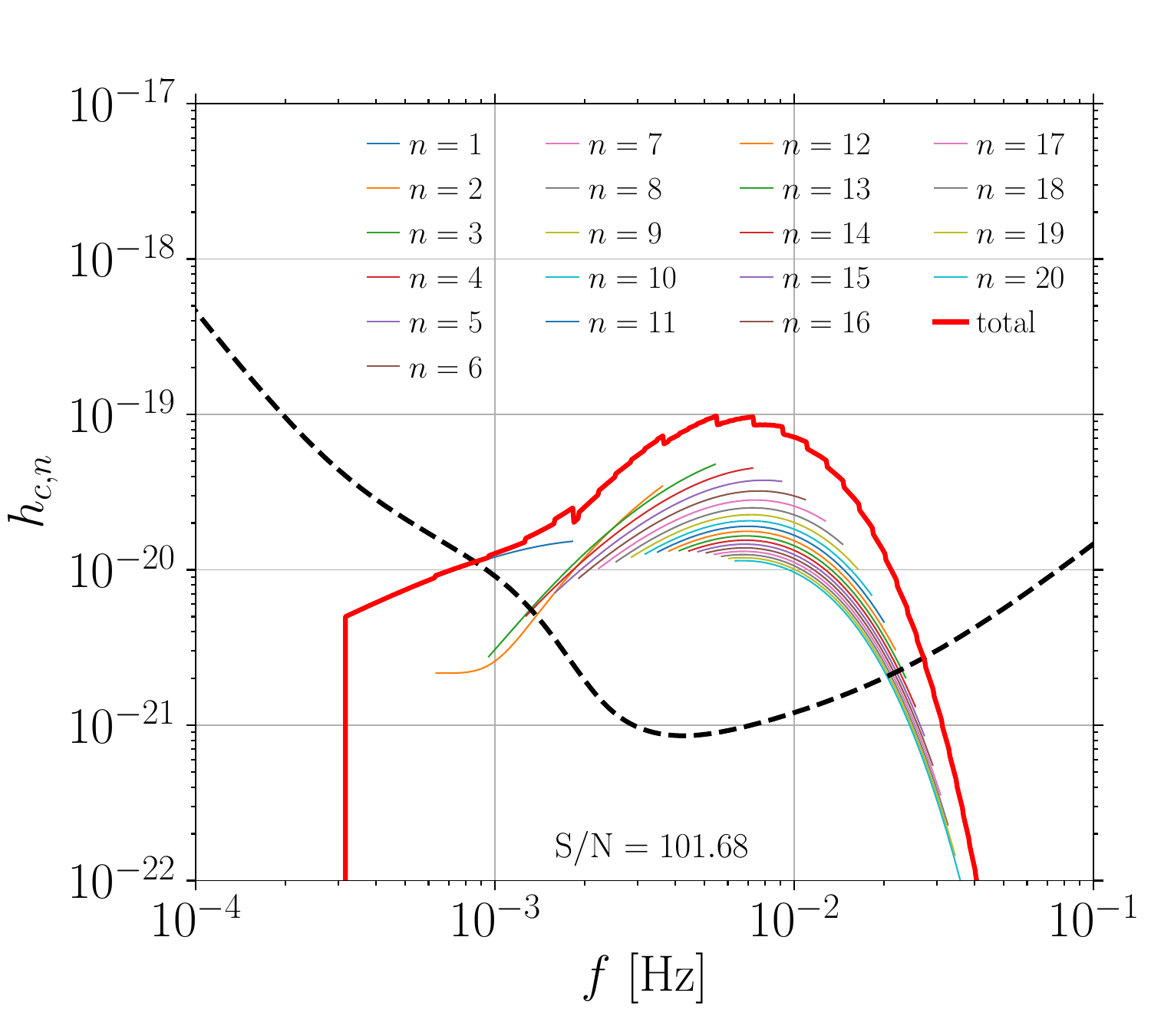}
    \includegraphics[width=0.48\textwidth]{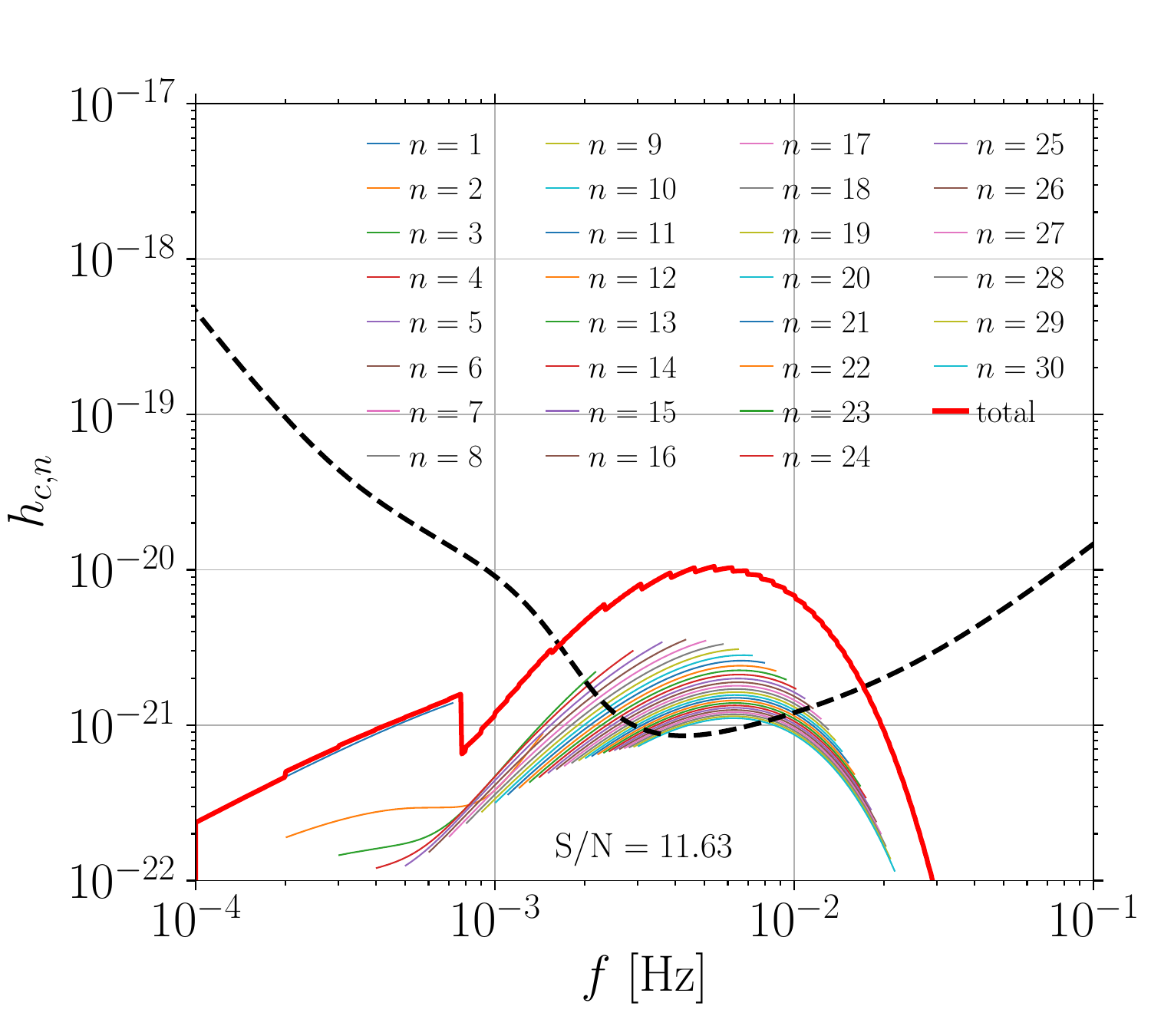}
    \caption{Waveform examples considering three different EMRI systems. {\it Upper left panel}: $z=1,m_1=10^5 ~{\rm M}_\odot, f_{\rm orb}/(1+z) = 10^{-3} ~{\rm Hz}, e = 0.5$. {\it Upper right panel}: $z=0.2,m_1=10^6 ~{\rm M}_\odot, f_{\rm orb}/(1+z) = 10^{-3.5} ~{\rm Hz}, e = 0.8$. {\it Bottom panel}: $z=2,m_1=10^6 ~{\rm M}_\odot, f_{\rm orb}/(1+z) = 10^{-4} ~{\rm Hz}, e = 0.9$. In each panel, the value of the estimated S/N is reported (assuming $T_{\rm obs} = 4$ yr and the LISA sensitivity curve of Section~\ref{sec:LISA_sens}). Note the different role played by high harmonics in the three different cases.}
    \label{fig:hc_example}
\end{figure*}

Formally, computing the sum $\Phi(f)$ in the above equation has the very same computational cost of evaluating equation~\eqref{eq:hcn} $n$ times, giving no advantages in the computation of equation~\eqref{eq:h_tot2}. Nevertheless, we note that at the leading Newtonian order in the GW back-reaction the eccentricity evolution turns out to be scale-free. In fact combining equation~\eqref{eq:dfdt_gw} and equation~\eqref{eq:dedt_gw} we get \citep[see e.g.][]{Enoki2007}

\begin{equation}\label{eq:e_f}
    \dfrac{f_{\rm orb}}{f_{\rm orb,0}}=\left[\dfrac{1-e_0^2}{1-e^2}\left(\dfrac{e}{e_0}\right)^{\frac{12}{19}}\left(\frac{1+121/304 e^2}{1+121/304 e_0^2}\right)^{\frac{870}{2299}}\right]^{-3/2},
\end{equation}
meaning that the eccentricity evolution is just a function of the frequency ratio $f_{\rm orb}/f_{\rm orb,0}$. As noted by \citet{Huerta2015}, this fact translates into a self-similar behaviour of $\Phi(f)$ when different initial eccentricities at fixed $f_{\rm obs,0}$ are selected. Specifically, they found that the location of the peak of $\Phi(f)$ (see Fig.~\ref{fig:Phi}) simply scales according to the following relation
\begin{equation}
  \frac{f_{p}}{f_{\rm orb,0}} =  \frac{1293}{181} \left[\frac{e_0^{12/19}}{1-e_0^2}\Bigl(1+\frac{121 e_0^2}{304}\Bigr)^{870/2299}\right]^{3/2}.
  \label{eq:fpeak}
\end{equation}
The spectrum of a binary with a different initial eccentricity $e_0$, specified at a different initial frequency $f_{\rm orb,0}$, can therefore be simply obtained by shifting the spectrum of a reference binary. In practice this consists in evaluating the function $\Phi(f)$ at 
\begin{equation}
    f = f' \dfrac{f_p}{f'_{p}} \dfrac{(1+z')}{(1+z)},
\end{equation}
where $f'_p$, $z'$ and $f'$ are the peak frequency, the redshift and the sampled frequency range of a reference binary. This scaling procedure results in a significant speed-up of the $h_c$ and S/N calculations,\footnote{The computational cost clearly depends on the number of harmonics taken into account. When such number is several thousand our procedure can up to 1-2 orders of magnitude faster since we avoid the cumbersome computation of $g_{n}(e_n)$ for each source.} making feasible the exploration of the EMRI GWB for several population models and under different assumptions. Ultimately, this motivates the adoption of the simple quadrupole formula for the EMRI evolution, instead of the more accurate PN formalism of BC04a.

The last point we need to address concerns the maximum and minimum orbital frequencies for which EMRIs are observed. Equation~\eqref{eq:e_f} formally holds in the frequency range $f_{\rm orb}\in(0,+\infty)$. However, at very low orbital frequencies the EMRI cannot be considered neither isolated from the dense stellar environment of the galactic nucleus nor GW-driven, while as the frequency increases due to the emission of gravitational radiation the CO will eventually plunge onto the MBH. Moreover, since LISA will observe EMRIs as transiting GW sources, that enter and coalesce inside its sensitivity window, it is particularly important to properly account for the finite duration of the emitted GW signal. Specifically, when computing the sum $\Phi(f)$ we need to discard the harmonics for which the orbital frequencies that produce contributions at a selected observed GW frequency $f$ lie outside the interval $[f_{\rm orb, min},f_{\rm orb, max}]$ and consider only those satisfying
\begin{equation}
    f_{\rm orb, min} \leq \dfrac{f(1+z)}{n} \leq f_{\rm orb, max}.
\end{equation}
The lowest frequency is simply settled by the orbital frequency of each EMRI at the beginning of the LISA observation run, while the highest one is determined by the maximum frequency reached at the end of the observation, which is $f_{\rm orb, max} = f_{\rm orb}(t=T_{\rm obs})$ if the EMRI does not plunge within $T_{\rm obs}$ or the ISCO frequency $f_{\rm orb, max} = c^3/(2\pi x^{3/2} G M)$, with $M$ the binary total mass and $x$ a factor multiplying the gravitational radius ($R_g = GM/c^2$) of the system, if the EMRI plunges within $T_{\rm obs}$. Assuming the primary MBH as non-spinning, this factor simply is $x=6$, while for the spinning case we assume that it can range in the interval $x \in [1,9]$ depending on the central MBH spin parameter $a$ and on the EMRI inclination.

In Fig.~\ref{fig:hc_example}, we report examples of characteristic strains for three eccentric sources together with the estimate of their S/N, evaluated as
\begin{equation}\label{eq:SNR_new}
    ({\rm S/N})^2 = \int \dfrac{h_{c}^2(f)}{f S_{\rm noise}(f)} \ud \ln f.
\end{equation}
with $h_{c}^2(f)$ given by equation~(\ref{eq:h_tot2}).
From the figure, we can infer that, depending on the initial orbital frequency and eccentricity, high harmonics can be quite relevant in shaping the total characteristic strain.

\subsection{EMRI catalogues}
\label{sec:catalogues}

\begin{table*}
\centering
\begin{tabular}{ccccccc|ccc}
\hline
 & Mass &  MBH & Cusp & $M$--$\sigma$ & & CO & \multicolumn{3}{c}{EMRI rate [$\mathrm{yr}^{-1}$]}\\
Model &  function &  spin &  erosion &  relation & $N_\mathrm{p}$ &  mass [$\msun$] & Total  & Detected (AKK)  & Detected (AKS) \\
\hline
M1 & Barausse12 & a98   & yes & Gultekin09    & 10  & 10 & 1600 &  294& 189  \\ 
M2 & Barausse12 & a98   & yes & KormendyHo13  & 10  & 10 & 1400 &  220& 146  \\ 
M3 & Barausse12 & a98   & yes & GrahamScott13 & 10  & 10 & 2770 &  809& 440  \\ 
M4 & Barausse12 & a98   & yes & Gultekin09    & 10  & 30 &  520 & 260 &221  \\ 
M5 & Gair10     & a98   & no  & Gultekin09    & 10  & 10 &  140 &  47& 15  \\ 
M6 & Barausse12 & a98   & no  & Gultekin09    & 10  & 10 & 2080 &  479& 261  \\ 
M7 & Barausse12 & a98   & yes & Gultekin09    & 0   & 10 & 15800 &  2712& 1765  \\ 
M8 & Barausse12 & a98   & yes & Gultekin09    & 100 & 10 &  180 &  35& 24  \\ 
M9 & Barausse12 & aflat & yes & Gultekin09    & 10  & 10 & 1530 &  217& 177  \\ 
M10 & Barausse12 & a0    & yes & Gultekin09    & 10  & 10 & 1520 &  188& 188 \\ 
M11 & Gair10     & a0    & no  & Gultekin09    & 100 & 10 &   13 &  1& 1 \\ 
M12 & Barausse12 & a98   & no  & Gultekin09    & 0   & 10 & 20000 &  4219& 2279 \\ 
\hline
\end{tabular}
\caption{
List of EMRI models taken from Babak17 and considered here to assess the GWB level. Column 1 defines the label of each model. For each model the following quantities are specified: the MBH mass function (column 2), the MBH spin model (column 3), whether or not the effect of cusp erosion is included (column 4), the $M$--$\sigma$ relation (column 5), the ratio of plunges to EMRIs (column 6), the mass of the COs (column 7), the total EMRI merger rate (yr$^{-1}$) up to $z=4.5$ (column 8). In column 9 and 10 the detected EMRI rate per year is reported for two different kind of waveforms (AKK and AKS, see Section~4 of Babak17 for full details) bracketing GW waveform modelling uncertainties.}
\label{tab:tab1}
\end{table*}

In order to provide astrophysical motivated estimates of the GWB generated by a cosmic population of EMRIs, we consider several models presented in Babak17, that are reported in Tab.~\ref{tab:tab1} for completeness. These models encompass a range of plausible prescriptions for the most relevant ingredients affecting EMRI formation, from the cosmic evolution of the MBH mass function to the relation between MBH mass and density of the surrounding stellar environment, from the rate of EMRI formation given the properties of the galactic nucleus to the occurrence ratio of direct plunges to EMRIs. We refer the reader to Babak17 for a detailed description of the underlying astrophysical models. For each EMRI population model, we obtained a catalog containing all EMRIs merging in the Universe (out to $z=4.5$) assuming 10 years of observation at Earth.

The GWB is likely generated not only by low S/N plunging systems (i.e with ${\rm S/N} < 20$, which is our standard detection threshold), but also by a large number of EMRIs emitting in the LISA band while still tens or even hundreds of years far from final plunge. By assuming a steady state evolution of the EMRI population, EMRIs plunging today have the same statistical properties of EMRIs plunging, say, 100 years from now. This means that EMRIs that are now 100 years from coalescence and are emitting GWs in the LISA band as we observe them during their slow adiabatic inspiral, have the same properties of EMRIs plunging now if they were observed 100 years ago. We can therefore derive the distribution of EMRIs far from coalescence in virtue of the continuity equation, simply by converting the merger rate $\ud N \ /\ud t$ (which we get from Babak17) into the population of emitting EMRIs sustaining that rate, i.e. $\ud N \ /\ud f=(\ud N \ /\ud t)(\ud t \ /\ud f)$.
We proceed as follows:
\begin{itemize}
    \item for each event in the catalog we draw the eccentricity at the last stable orbit, $e_p$, from a flat distribution in the range $[0,0.2]$, thus obtaining all the relevant properties of the event: $(M,z,e_{\rm p})$;
    \item we integrate the orbital elements of the event backward in time for $T_{\rm back}$ years;
    \item we then randomly sample $N_{\rm back} = {\rm int} (T_{\rm back}/10)$ points in the range $[0,T_{\rm back}]$ in order to select different evolutionary points of a specific EMRI;\footnote{The division by 10 is due to the fact that we collect 10 catalogues of EMRIs coalescence, each of which is meant to represent one year of observation.}
    \item we record the orbital frequency and eccentricity for each of $N_{\rm back}$ points, that effectively will represent new EMRIs to be then evolved for the duration of LISA mission.
\end{itemize}
With the above procedure we can effectively build-up a proxy of the whole population of EMRIs that, as implied by the continuity equation, guarantees the merger rate observed in the synthetic catalogues. This corresponds to formally transform the differential EMRI merger rate $\ud^4 N/(\ud z \ud M \ud e_p \ud t)$ into a differential EMRI inspiral rate $\ud^4 N/(\ud z \ud M \ud f \ud e_f)$, where $f$ and $e_f$ are self-consistently evaluated by numerically integrating equation~\eqref{eq:dfdt_gw} and equation~\eqref{eq:dedt_gw} from $T_{\rm back}$ years prior to the plunge. In order to optimise our sampling we adopt a $T_{\rm back}$ that depends on the MBH mass, specifically we assume
\begin{equation}
    T_{\rm back} = 20 \left(\dfrac{m_1}{10^4 \rm M_\odot}\right) \rm yr.
\end{equation}
Our choice is determined by the fact that the time to cover the same range in gravitational radii scales linearly with the mass of the system, thus for more massive MBHs the GW dominated evolution takes longer from a fixed initial separation (in units of $GM/c^2$) down to the last stable orbit of the system.
Practically, this means that EMRIs orbiting around low mass MBHs emit in the LISA band only over the last few years of their evolution, whereas those orbiting high mass MBHs emit in the LISA band already several hundreds of years prior to final plunge. 

A caveat of the above procedure is that, practically, we are building a population made of several copies of the coalescing EMRIs acquired from the catalogues of Babak17. In particular for each EMRI there will be $N_{\rm back}$ other copies with same redshift and primary mass, rather than a completely independent population. Despite this limitation, the 10 year EMRI catalogues we are using contain several thousands of events covering the relevant MBH mass and redshift range. Therefore, we do not expect our ``copying procedure'' to introduce any significant bias in the GWB generation process.

\subsection{Background computation}
\label{sec:gwb_comput}

Using the source catalogues generated with the procedure described in the previous subsection, we are now in the position to evaluate the GWB generated by EMRIs. To this end, we start by providing a brief description of the formalism employed for the evaluation of such GWB, generally valid for any a cosmic population of GW-driven binaries, not necessarily EMRIs.

Following \citet{Phinney2001}, the characteristic strain of the GWB generated by an inspiralling cosmic population of sources can be expressed in terms of their comoving number density ($n_{c}$) combined with the emitted GW energy spectrum ($\ud E/ \ud \ln f_r$), i.e:

\begin{widetext}
\begin{align}\label{eq:hc_P}
    h_{c,\rm gwb}^2(f) = \dfrac{4 G}{\pi c^2 f^2} \int \ud z \ud \mathcal{M} \dfrac{\ud^2 n_{c}}{\ud z \ud \mathcal{M}} \dfrac{1}{1+z} \dfrac{\ud E}{\ud \ln f_r}\Biggl|_{f_r = f(1+z)}.
\end{align}
\end{widetext}
The comoving density can be turned into a merger rate, i.e. the number ($N$) of sources merging per unit time,
\begin{align}
    \dfrac{\ud^2 n_c}{\ud z \ud \mathcal{M}} &= \dfrac{\ud^3 N}{\ud z \ud \mathcal{M} \ud t_r} \dfrac{1}{(1+z) 4\pi c d^2},
\end{align}
that, with additional manipulations, can be further related to the frequency distribution of sources, i.e. 
\begin{align}\label{eq:dN}
    \dfrac{\ud^2 n_c}{\ud z \ud \mathcal{M}} &=  \dfrac{1}{(1+z) 4\pi c d^2} \dfrac{\ud^3 N}{\ud z \ud \mathcal{M} \ud \ln f_{\rm orb}} \dfrac{\ud \ln f_{\rm orb}}{\ud t_r}.
\end{align}

Turning to the energy spectrum of GW-driven binaries (as detailed in Section~\ref{sec:SNR}), we know that when the eccentricity is non-zero the GW power is emitted at several harmonics of the orbital motion. The spectrum can be therefore expressed as a sum of contributions coming from each harmonic

\begin{align}\label{eq:dE}
    \dfrac{\ud E(f_r)}{\ud \ln f_r} &= f_r \dfrac{\ud E}{\ud t_r}\dfrac{\ud t_r}{\ud f_r}\nonumber\\
    &= \sum_n \dfrac{\ud E_n}{\ud t_r} f_r \dfrac{\ud t_r}{\ud f_r}\nonumber\\
    & = \sum_n \dfrac{\ud E_n}{\ud t_r} n\dfrac{f_r}{n} \dfrac{\ud t_r}{\ud (n f_r/ n)}\nonumber\\
     & = \sum_n \dfrac{\ud E_n}{\ud t_r} \dfrac{\ud t_r}{\ud \ln f_{\rm orb}}
\end{align}
where $f_{\rm orb} = f_r/n$ is assumed to hold.
In the above expressions there is also an implicit dependence on the orbital eccentricity that strongly influences the residence time at a particular frequency. We therefore need to take into account the initial eccentricity distribution of sources to correctly estimate the GWB. Inserting equation~\eqref{eq:dN} and equation~\eqref{eq:dE} in equation~\eqref{eq:hc_P} we get

\begin{align}
    h_{c,\rm gwb}^2(f) = &\int \ud z \ud \mathcal{M} \ud e \Biggl[ \sum_n \dfrac{\ud^4 N}{\ud z \ud \mathcal{M} \ud e \ud \ln f_{\rm orb}}\times \nonumber\\
    &\times \dfrac{G \dot{E}_n}{(1+z)^2 \pi^2 c^3 d^2 f^2} \Biggr]_{f_{\rm orb} = \frac{f (1+z)}{n}}.
\end{align}
Substituting $f = n f_{\rm orb}/(1+z)$ and recalling that the rms strain of the $n$-th harmonic is defined by \citep{Finn2000,Amaro-Seoane2010} 
\begin{equation}\label{eq:rms_strain}
    h_n^2 = \dfrac{G \dot{E}_n}{c^3 \pi^2 d^2 f_n^2},\\
\end{equation}
we obtain

\begin{align}\label{eq:hc_sum}
    h_{c,\rm gwb}^2(f) = &\int \ud z \ud \mathcal{M} \ud e \ \times \nonumber\\
    &\times \Biggl[ \sum_n \dfrac{\ud^4 N}{\ud z \ud \mathcal{M} \ud e \ud \ln f_{\rm orb}} h_n^2(f) \Biggr]_{f_{\rm orb} = \frac{f (1+z)}{n} }
\end{align}
where, at each observed frequency $f$ and for each $n$ in the sum, only binaries with the proper orbital frequency (i.e. $f_{\rm orb} = f (1+z)/n$) contribute to $h_{c,\rm gwb}^2$. Note that in the particular case of a population of circular binaries the above equation reduces to the well known form worked out by, e.g. \citet{Sesana2008}.

Equation~\eqref{eq:hc_sum} assumes that binaries do not evolve significantly under GW back-reaction during the observation time span $T_{\rm obs}$. Despite this does not represent a problem for some class of sources, e.g. massive black hole binaries (MBHBs) in the PTA band, for EMRIs this approximation turns out to be unrealistic. In fact, during typical $T_{\rm obs}$ of months to years, each single EMRI harmonic can span a substantial frequency range, as shown by the tracks visualized in Fig.~\ref{fig:hc_example}.

In this case, we can adapt equation~\eqref{eq:hc_sum} by weighting the contribution of each harmonic with the ratio of the number of cycles spent by the source at a given frequency (i.e. $f^2/\dot{f}$) and the maximum number of cycles observable at that frequency for a given observation duration $T_{\rm obs}$, i.e.:

\begin{equation}
    \label{eq:cyc}
    \dfrac{N_{\rm cyc, gw}}{N_{\rm cyc, obs}} = f \dfrac{\ud t}{\ud \ln f} \times \dfrac{1}{f T_{\rm obs}}.
\end{equation}
Note that since observations are limited by $T_{\rm obs}$, this ratio cannot be higher than one, so when $T_{\rm obs}<f/\dot{f}$ we go back to the non-evolving case discussed above.
Then recalling that in equation~\eqref{eq:hc_sum} $f_{\rm orb} = f (1+z)/n$ must hold, we can write

\begin{align}
   &h_n^2 \times f \dfrac{\ud t}{\ud \ln f} \times \dfrac{1}{f T_{\rm obs}} \nonumber\\
   &= h_n^2 \times \dfrac{n f_{\rm orb}}{1+z} \dfrac{\ud t}{\ud \ln (n f_{\rm orb}/(1+z))} \times \dfrac{1}{f T_{\rm obs}} \nonumber\\
   &= h_n^2 \times \dfrac{\ud t_r}{\ud \ln (n f_{\rm orb})} n f_{\rm orb} \times \dfrac{1}{f T_{\rm obs}}\nonumber\\
   &= h_n^2 \times \dfrac{\ud t_r}{\ud \ln f_n} f_n \times \dfrac{1}{f T_{\rm obs}}\nonumber\\
   &= \dfrac{1}{2} h_{c,n}^2 \times \dfrac{1}{f T_{\rm obs}}
   \label{eq:hcnevol}
\end{align}
where in the last line we used the definition of the characteristic strain of the $n$-th harmonic \citep[compare equation~\ref{eq:hcn} and equation~\ref{eq:rms_strain} and check with Ref.][]{Finn2000}. 
Finally, we can modify equation~\eqref{eq:hc_sum} as

\begin{align}\label{eq:hc_evo}
    &h_{c,\rm gwb}^2(f) = \dfrac{1}{2}\int \ud z \ud \mathcal{M} \ud e \Biggl[ \sum_n \dfrac{\ud^4 N}{\ud z \ud \mathcal{M} \ud e \ud \ln f_{\rm orb}} \times \nonumber\\
    & \times \dfrac{h_{c,n}^2(f)}{f T_{\rm obs}} \Biggr]_{f_{\rm orb} = \frac{f (1+z)}{n} \in [f_{\rm orb, min},f_{\rm orb, max}]}
\end{align}
where we folded in the additional requirement $f_{\rm orb} \in [f_{\rm min},f_{\rm max}]$, with extrema defined as

\begin{align}
    f_{\rm orb, min} &= f_{\rm orb}(t=0),\nonumber\\
    f_{\rm orb, max} &= \text{min}\Bigl(f_{\rm orb, ISCO},f_{\rm orb}(t=T_{\rm obs})\Bigr),
\end{align}
to account for the frequency evolution of sources during $T_{\rm obs}$. Practically, this condition selects the GW radiation emitted within $T_{\rm obs}$, i.e. in the time interval in which the orbital frequency changes from $f_{\rm orb}(t=0)$ (the orbital frequency at beginning of the observation) to $f_{\rm orb, max}$ (the maximum orbital frequency reached, either at the end of the observation period or at binary coalescence). The above condition can be also recast to explicitly select a limited harmonic number range. In fact for a given observed GW frequency $f$ we require, as said, that $f_{\rm orb} = f(1+z)/n$ must hold. The fact that $f_{\rm orb}$ has to lie between a minimum an maximum value translates into requiring that the summation in equation~\eqref{eq:hc_evo} spans the harmonic range $[n_{\rm min}, n_{\rm max}]$, where 
\begin{align}
    n_{\rm min} &= \dfrac{f (1+z)}{ f_{\rm orb, max}},\nonumber\\
    n_{\rm max} &= \dfrac{f (1+z)}{ f_{\rm orb, min}},
\end{align}
in which it is implicit that only the integer part of the right hand side has to be considered. Equation~\eqref{eq:hc_evo} therefore reads

\begin{widetext}
\begin{align}\label{eq:hc_evo_n}
    h_{c,\rm gwb}^2(f) = & \dfrac{1}{2} \int \ud z \ud \mathcal{M} \ud e
   \Biggl[ \sum_{n=n_{\rm min}}^{n_{\rm max}} \dfrac{\ud^4 N}{\ud z \ud \mathcal{M} \ud e \ud \ln f_{\rm orb}} \dfrac{h_{c,n}^2(f)}{f T_{\rm obs}} \Biggr]_{f_{\rm orb} = \frac{f (1+z)}{n}}
\end{align}
\end{widetext}

The practical computation of the GWB from equation~\eqref{eq:hc_evo} proceeds as follows. The distribution $\ud^4 N/(\ud z \ud \mathcal{M} \ud e \ud \ln f_{\rm orb})$ is in fact a finite list of sources computed following the procedure described in Section~\ref{sec:catalogues}. This turns all the integrals in equation~\eqref{eq:hc_evo} into sums over the list of sources. For each EMRI $f_{\rm orb}$ and $e$ are defined at the start of the LISA mission $t=0$. The EMRI orbital elements are then integrated between $t=0$ and $t=T_{\rm obs}$ using equation~\eqref{eq:dfdt_gw} and equation~\eqref{eq:dedt_gw} to obtain the range $f_{\rm orb, min} < f_{\rm orb} < f_{\rm orb, max}$ and the function $e(f_{\rm orb})$ to be considered when evaluating its contribution to the GWB. If we are computing the GWB at frequency $f$, we take all the harmonics for which $f(1+z)/n \in [f_{\rm orb, min},f_{\rm orb, max}]$ is satisfied and we add to $h_{c,\rm gwb}^2(f)$ a contribution $h_{c}^2(f)/(fT_{\rm obs})$, where $h_{c}(f)$ is given by equation~\eqref{eq:h_tot2}, with $\Phi(f)$ limited between $[n_{\rm min},n_{\rm max}]$, while the eccentricity is evaluated using the pre-computed $e(f)$ evolution. Note that since $h_{c}$ does not depend on $T_{\rm obs}$, the contribution of each single harmonic of each individual EMRI to the GWB, $h_{c,n}^2(f)/(fT_{\rm obs})$, is inversely proportional to $T_{\rm obs}$. However, as $T_{\rm obs}$ increases, so does the frequency range, $f_{\rm orb, min} < f < f_{\rm orb, max}$, to which each individual EMRI contributes. This means that, the number of EMRIs contributing at a given $f$ increases proportionally to $T_{\rm obs}$. Therefore, ignoring subtleties related to the resolvability of individual sources, the signal computed via equation~\eqref{eq:hc_evo} is independent on $T_{\rm obs}$, as expected for an unresolved GWB.

Finally, the detectability of the GWB is assessed by computing the associated power signal-to-noise ratio (S/N$_{\rm gwb}$) through \citep{Thrane2013,2016PhRvL.116w1102S}
\begin{equation}\label{eq:power_SNR}
    ({\rm S/N_{\rm gwb}})^2 = T_{\rm obs} \int \gamma(f) \dfrac{h_{c,\rm gwb}^4(f)}{f^2 S_{\rm noise}^2(f)} \ud f
\end{equation}
where $h^2_{c, \rm gwb}(f)$ is the characteristic strain of the GWB, $S_{\rm noise}(f)$ is the power spectral density of LISA (see Section~\ref{sec:LISA_sens}), while the function $\gamma(f)$ is assumed to be approximately constant and equal to unity \citep[see Fig.~4 in Ref.][]{Thrane2013}. We caution that we do not consider any WD confusion noise subtraction when computing the EMRI GWB S/N via equation~\eqref{eq:power_SNR}.

\section{Results}
\label{sec:results}

In this section, we present the main results of our investigation, including i) the detectability of the EMRI GWB and its S/N computed via equation~\eqref{eq:power_SNR} and, ii) the number of individually resolvable sources. Relevant numbers are computed as a function of $T_{\rm obs}$ for different astrophysical EMRI models and under a variety of assumption for the EMRI waveform model as well as for the GWB computation, as described below. Tab.~\ref{tab:tab2} quantifies our main results for three representative models and will serve as a pivotal tool for discussion in the following sub-sections. 

\subsection{Expected EMRI GWB: general considerations}
\label{sec:3.1}

\begin{figure*}
    \centering
    \includegraphics[width=0.48\textwidth]{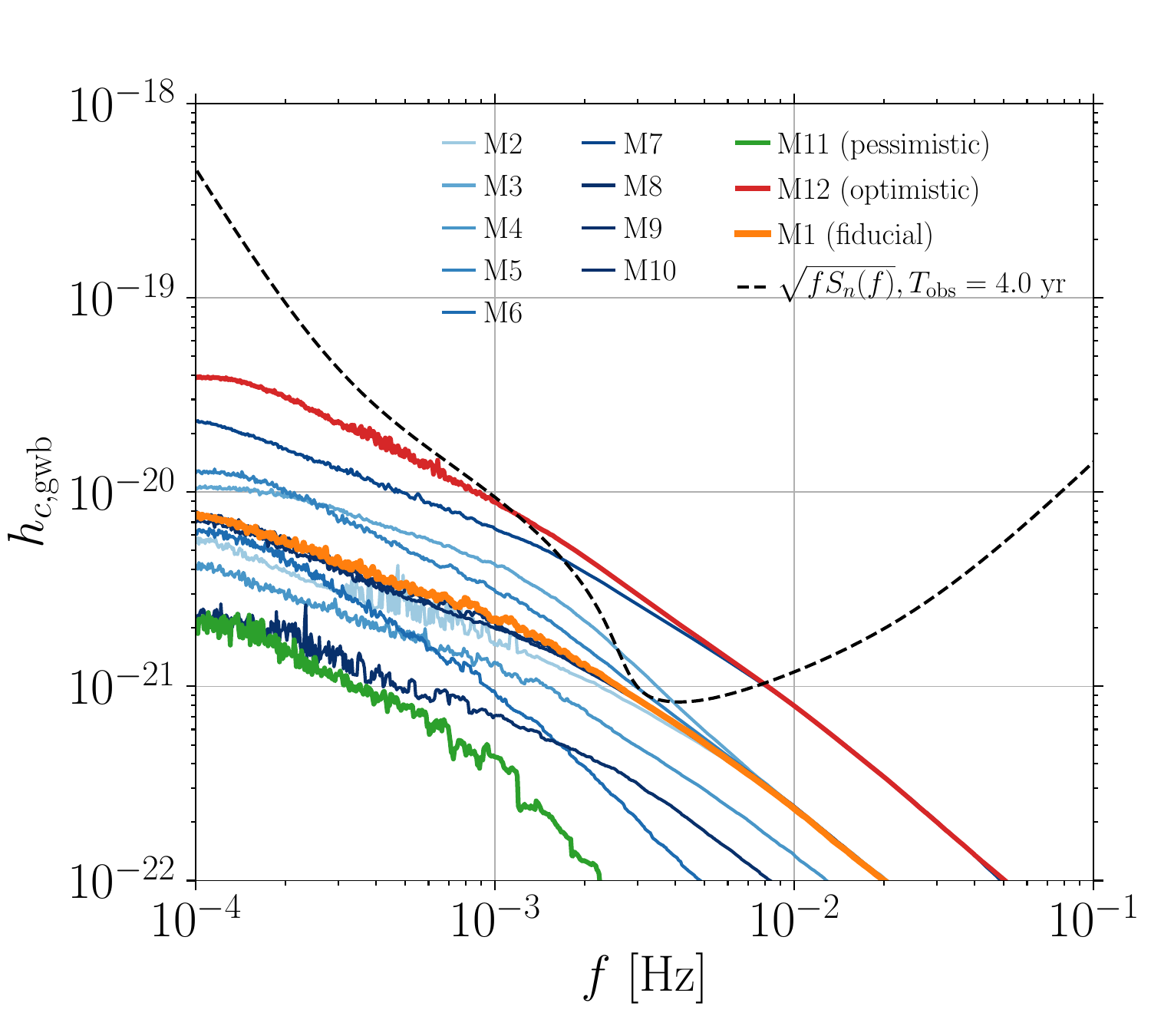}
    \includegraphics[width=0.48\textwidth]{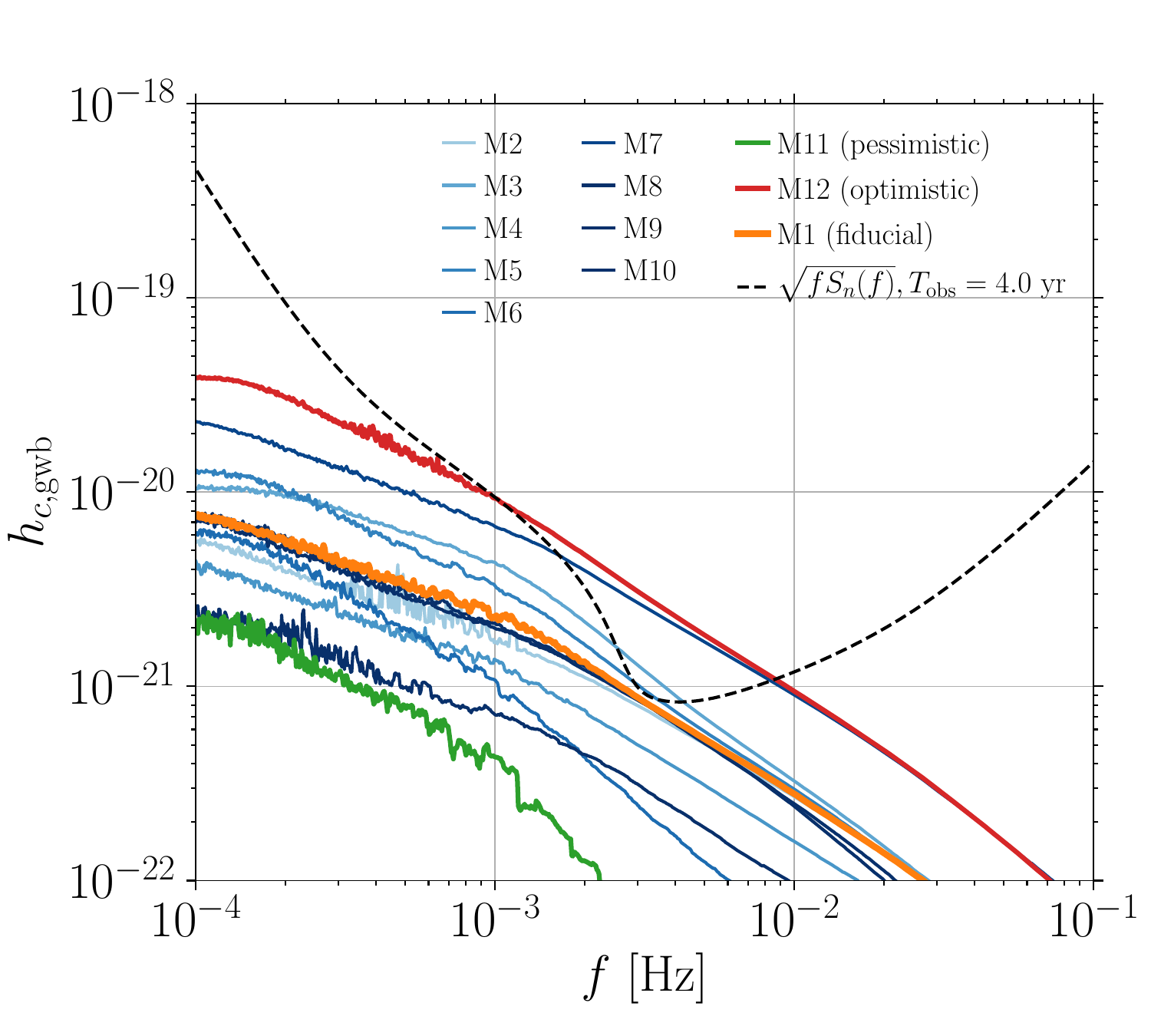}
    \caption{Characteristic strain of the GWB generated in the 12 different EMRI formation scenarios reported in Tab.~\ref{tab:tab1}. {\it Left panel}: last stable orbit is chosen at $6 G M/c^2$. {\it Right panel}: the last stable orbit varies according to the spin of $m_1$.}
    \label{fig:gwb_all1}
\end{figure*} 
In Fig.~\ref{fig:gwb_all1} we show the characteristic GWB strain predicted by the 12 models of Tab.~\ref{tab:tab1} compared to the LISA sensitivity curve (black dashed line). Given the significant uncertainties in the EMRI waveform modeling close to the final plunge, following Babak17, we ran two different sets of models, truncating the EMRI GW signal either at the Schwarzschild (analytic kludge Schwarzschild, AKS, waveform model) or at the Kerr ISCO (analytic kludge Kerr, AKK, waveform model). In the AKS case, shown in the left panel of Fig.~\ref{fig:gwb_all1}, all EMRIs plunge at $R_{\rm ISCO}=6 G M/c^2 = 6 R_g$. In the AKK case, shown in the right panel of Fig.~\ref{fig:gwb_all1}, we use the MBH spin information recorded in our MBH population synthesis model to identify the last stable orbit in the range $1<R_{\rm ISCO}/R_g< 9$, depending on the MBH spin and EMRI orbital inclination. In this way, co-rotating EMRIs remain stable much closer to the horizon, while counter-rotating ones become unstable and start their radial infall at farther distance. This, in principle, could make a substantial difference in the signal computation, since our population synthesis model predicts that MBHs producing EMRIs in the LISA range tend to be highly spinning. However, this does not seem to be the case and, regardless of the ISCO assumption, we find that two thirds (8 out of 12) of the investigated models result in a GWB with $h_{c, \rm gwb}$ comparable to or higher then the LISA noise curve. 

To ease the discussion, we focus our attention on the two models bracketing the GWB uncertainty range, i.e. the pessimistic model M11 (green line) and the optimistic model M12 (red line) and the fiducial model M1 (orange line), which lies half way between them, grazing the LISA sensitivity curve close to the bucket. Uncertainties in the EMRI GWB estimate span about 1.5 orders of magnitude in $h_c$, which is consistent with the three orders of magnitude uncertainty in the EMRI rates reported by Babak17. Two things are worth noticing. First, a large fraction of the investigated models, including the fiducial M1, predicts a GWB comparable to or higher than the LISA sensitivity curve, which therefore cannot be neglected when considering detectability of other sources. Second, the more optimistic scenario M12 can be a blessing for EMRIs but a curse for other sources. In fact, in this case, the EMRI GWB would deteriorate LISA sensitivity in the bucket by more than a factor of three. 

\begin{figure}
    \centering
    \includegraphics[width=0.48\textwidth]{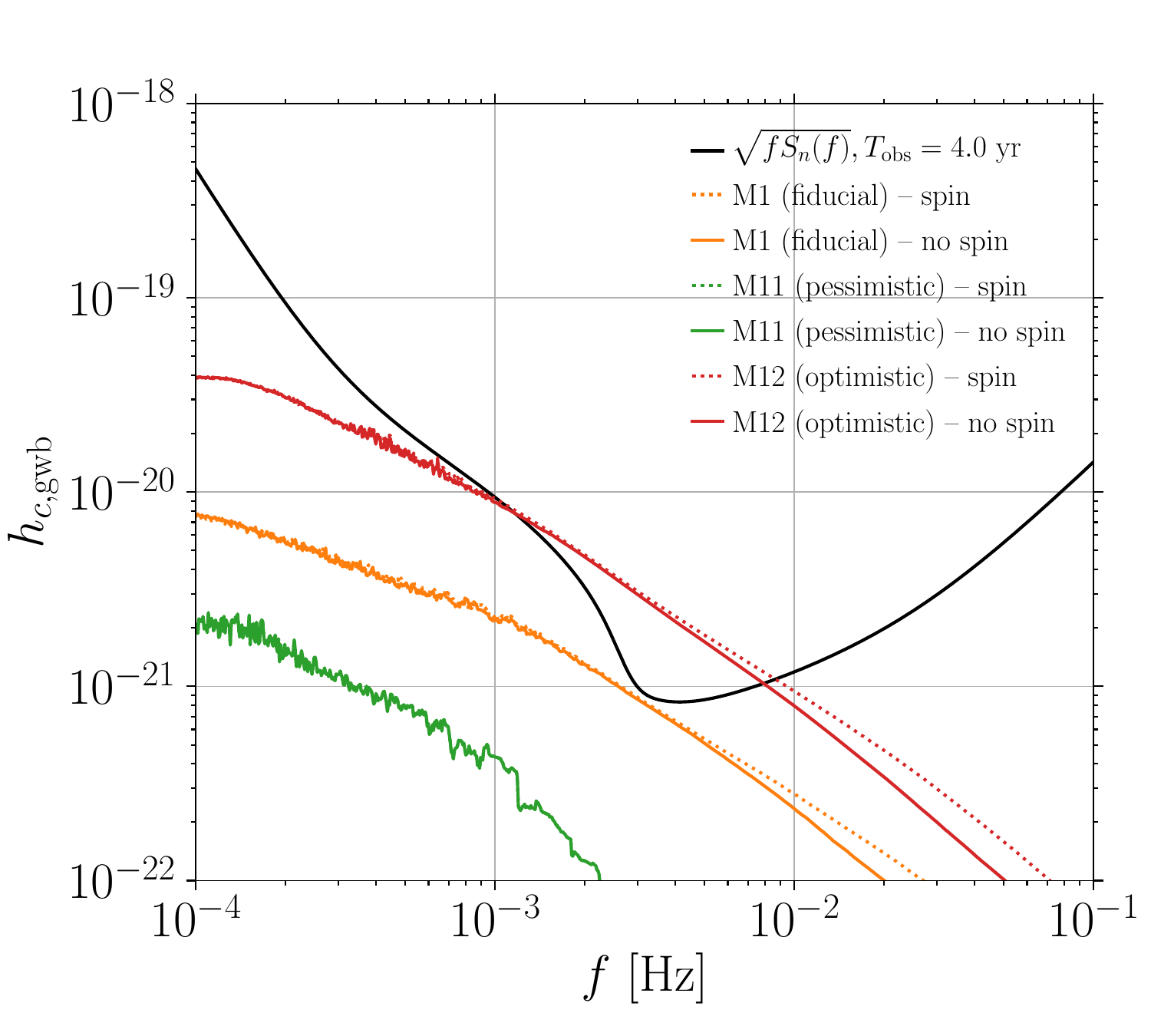}
    \caption{Characteristic strain of the GWB for the fiducial and extreme models when the last stable orbit depends on the spin of $m_1$ (dotted lines) or not (solid lines). Noticeable differences arise only above $\approx 10^{-2}$ Hz, where in the spinning case the GWB results to be slightly higher.}
    \label{fig:gwb_spin_nospin}
\end{figure}

A detailed comparison of the dependence of the signal on the location of the ISCO for M1, M11 and M12 is shown in Fig.~\ref{fig:gwb_spin_nospin}. 
In practice, the ISCO location only matters above $f \approx 10^{-2}$ Hz, where in the spinning case (AKK) the GWB is slightly higher. This is the result of two competing effects. On the one hand, when adopting the Kerr ISCO (AKK), the CO can get deeper within the MBH gravitational potential before eventually plunging, thus emitting more power and at higher frequencies. On the other hand, this also means that EMRIs have higher S/N on average, and more of them can be individually resolved and subtracted from the GWB. The two effects almost cancel but the latter is subdominant and the resulting GWB is slightly higher. This is reflected in the GWB S/N reported in Tab.~\ref{tab:tab2}, where a comparison between the AKK and AKS columns shows that the waveform choice has a mere 5\% impact on the S/N of the GWB. Reported numbers certify that the EMRI GWB is easily detectable in models M1 and M12, with S/N of several hundreds and several thousands respectively already after few months of data collection. Even in the most pessimistic case (M11), despite the signal level is an order of magnitude below the detector sensitivity (cf. green curve in Fig.~\ref{fig:gwb_spin_nospin}), the power S/N is of the order of a few and the signal might be marginally detected within few years of observation.

\begin{table}
    \centering
    \begin{tabular}{cc|ccc|ccc}
    \hline
    
    \multirow{2}{*}{Model} & \multirow{2}{*}{$T_{\rm obs}$} & \multicolumn{3}{c|}{detections} & \multicolumn{3}{c}{S/N$_{\rm gwb}$}\\
    & & AKK & AKS & AKSb & AKK & AKS & AKSb \\ 
    \hline
    
    \multirow{5}{*}{M1} & 0.5   yr & 23   & 10   & 9   & 283 & 273 & 273 \\
                        & 1.0   yr & 61   & 39   & 27  & 361 & 347 & 354 \\
                        & 2.0   yr & 166  & 107  & 80  & 461 & 446 & 457 \\
                        & 4.0   yr & 466  & 372  & 234 & 569 & 536 & 582 \\
                        & 10.0  yr & 1586 & 1267 & 827 & 747 & 711 & 793 \\
    \hline
    
    \multirow{5}{*}{M11} & 0.5   yr & 0 & 0 & 0 & 2.3 & 2.3 & 2.3 \\
                         & 1.0   yr & 0 & 0 & 0 & 2.7 & 2.7 & 2.7 \\
                         & 2.0   yr & 0 & 0 & 0 & 3.6 & 3.6 & 3.6 \\
                         & 4.0   yr & 0 & 0 & 0 & 5.0 & 5.0 & 5.0 \\
                         & 10.0  yr & 1 & 1 & 1 & 6.8 & 6.8 & 6.8 \\
    \hline
    
    \multirow{5}{*}{M12} & 0.5  yr & 331   & 194   & 57   & 3907 & 3327  & 3581  \\
                         & 1.0  yr & 905   & 521   & 122  & 4805 & 4253  & 4686  \\
                         & 2.0  yr & 2444  & 1515  & 284  & 5936 & 5301  & 6231  \\
                         & 4.0  yr & 6492  & 4126  & 690  & 7166 & 6582  & 8297  \\
                         & 10.0 yr & 20698 & 13551 & 2085 & 9424 & 8849  & 12279 \\
    \hline
    \end{tabular}
    \caption{Number of resolvable sources and residual GWB S/N for models M1, M11 and M12 as a function of the mission duration. Reported numbers represent: mission duration (column 2) number of individually resolvable  sources (columns 3, 4 and 5) and S/N of the GWB (columns 5, 6 and 7). Results are shown for both the AKK and AKS waveforms. Finally, results marked as AKSb are obtained considering the LISA curve augmented by the corresponding EMRI GWB (i.e. considering it as an additional noise source, see Fig.~\ref{fig:gwb_all2} and main text for details).}
    \label{tab:tab2}
\end{table}

In any case, the difference is small and practically irrelevant in the LISA band. Therefore, pending potential issues related to accurate waveform modeling for signal identification, we conclude that the MBH spin distribution and the details of the signal close to the final plunge do not appreciably affect the level of the EMRI GWB. With this understanding, in the following, we present results for the AKS models only.

\begin{figure}
    \centering
    \includegraphics[width=0.48\textwidth]{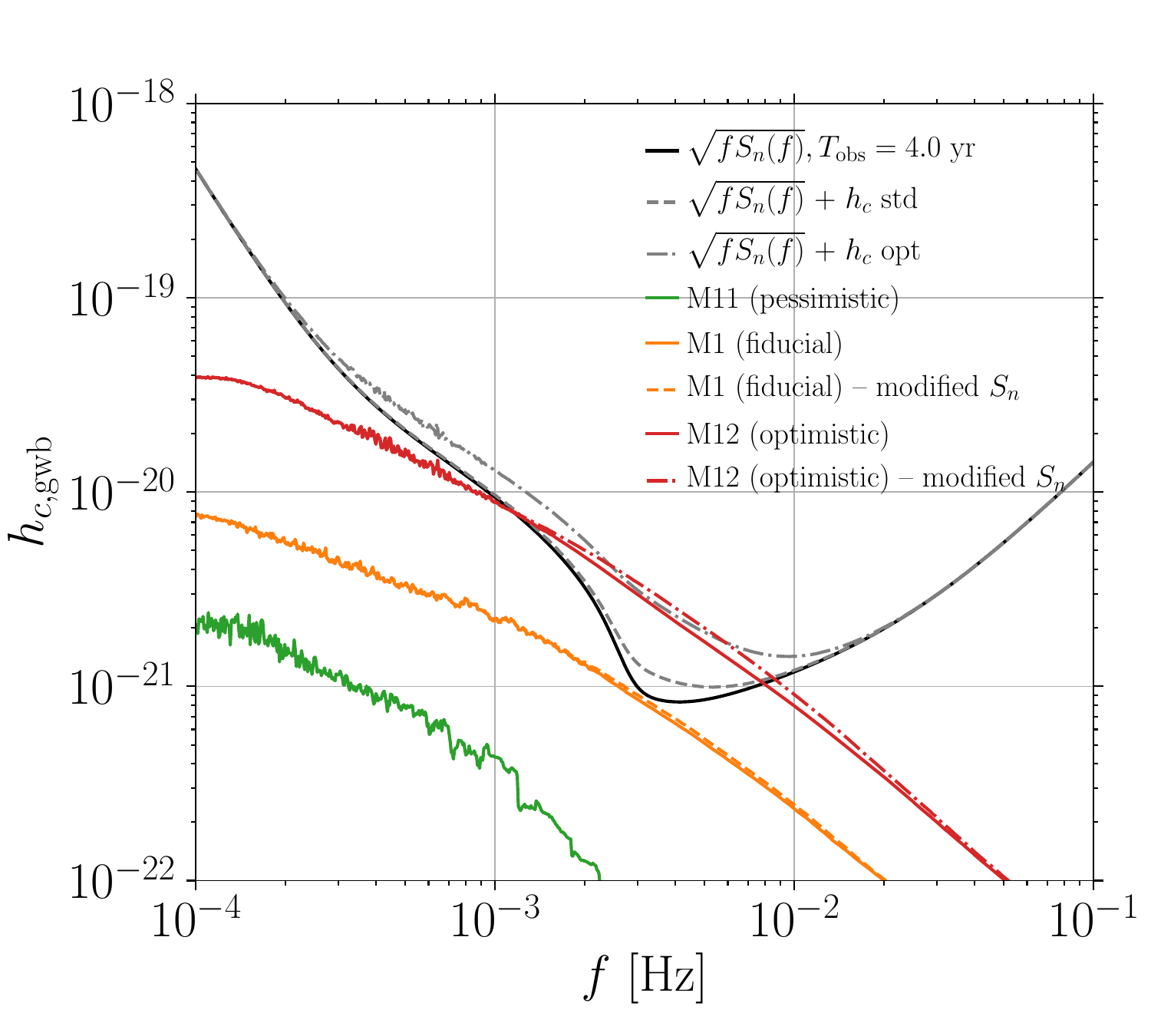}
    \caption{Characteristic strains of the GWB generated in the fiducial model (orange lines) and those of the models that bracket from the lower (pessimistic, green line) and the upper branch (optimistic, red line) all the investigated EMRI formation scenarios. Dashed (M1) and dashed-dotted (M12) lines denote the EMRI GWB and the LISA noise curve (grey) when, for each model, the associated GWB is heuristically integrated in the LISA sensitivity as an additional noise source (see text).}
    \label{fig:gwb_all2}
\end{figure}

Results shown so far were obtained taking into account LISA instrumental noise and WD confusion noise only when computing individual EMRI S/N and subtracting resolvable signals. In practice, this procedure is bound to likely underestimate the resulting GWB, since the GWB itself should be taken into account when computing individual EMRI S/N. A rigorous estimate of the GWB should therefore be done by subtracting resolvable sources one by one while including the overall signal produced by other systems. This is expected to lower the S/N of individual sources leaving behind a larger GWB. To bracket uncertainties due to our simplistic procedure, we also ran a set of models adding in quadrature to the instrument noise the EMRI GWB previously estimated by using the LISA noise only. Results are shown in Fig.~\ref{fig:gwb_all2} for the test cases M1, M11 and M12.
The overall resulting LISA sensitivity is shown by the grey dashed and dotted-dashed curves for models M1 and M12 respectively. For M1 we observe that the sensitivity curve is shifted upward by about 30-40\% in the bucket, while for M12 the sensitivity gets dramatically affected with upward shifts up to a factor of $\approx4$. As expected, the corresponding GWB for M1 and M12 evaluated with these degraded sensitivity curves is slightly higher. Differences are contained within 20-25\% in the worst case scenario (M12), thus certifying that our simple GWB amplitude estimates are robust. These figures are reflected in the S/N reported in Tab.~\ref{tab:tab2} (columns AKS vs AKSb): when the GWB is included in the computation, S/N are generally few \% larger for M1, up to 20\% larger for M12 and completely unaffected for M11. We notice, however, that even a small change in the GWB amplitude can have an important impact on the number of resolvable sources, especially because EMRI detection is S/N threshold limited. For example for M1 we obtain 234 (372) individually resolvable EMRIs in $T_{\rm obs}=4\,$ yr when the underlying EMRI GWB is (is not) taken into account in their S/N evaluation. The difference becomes even more striking in the M12 models, with individual EMRI detections dropping by a factor of seven, from 4126 to 690.

\subsection{The build-up of the EMRI GWB}

\begin{figure}
    \centering
    \includegraphics[width=0.48\textwidth]{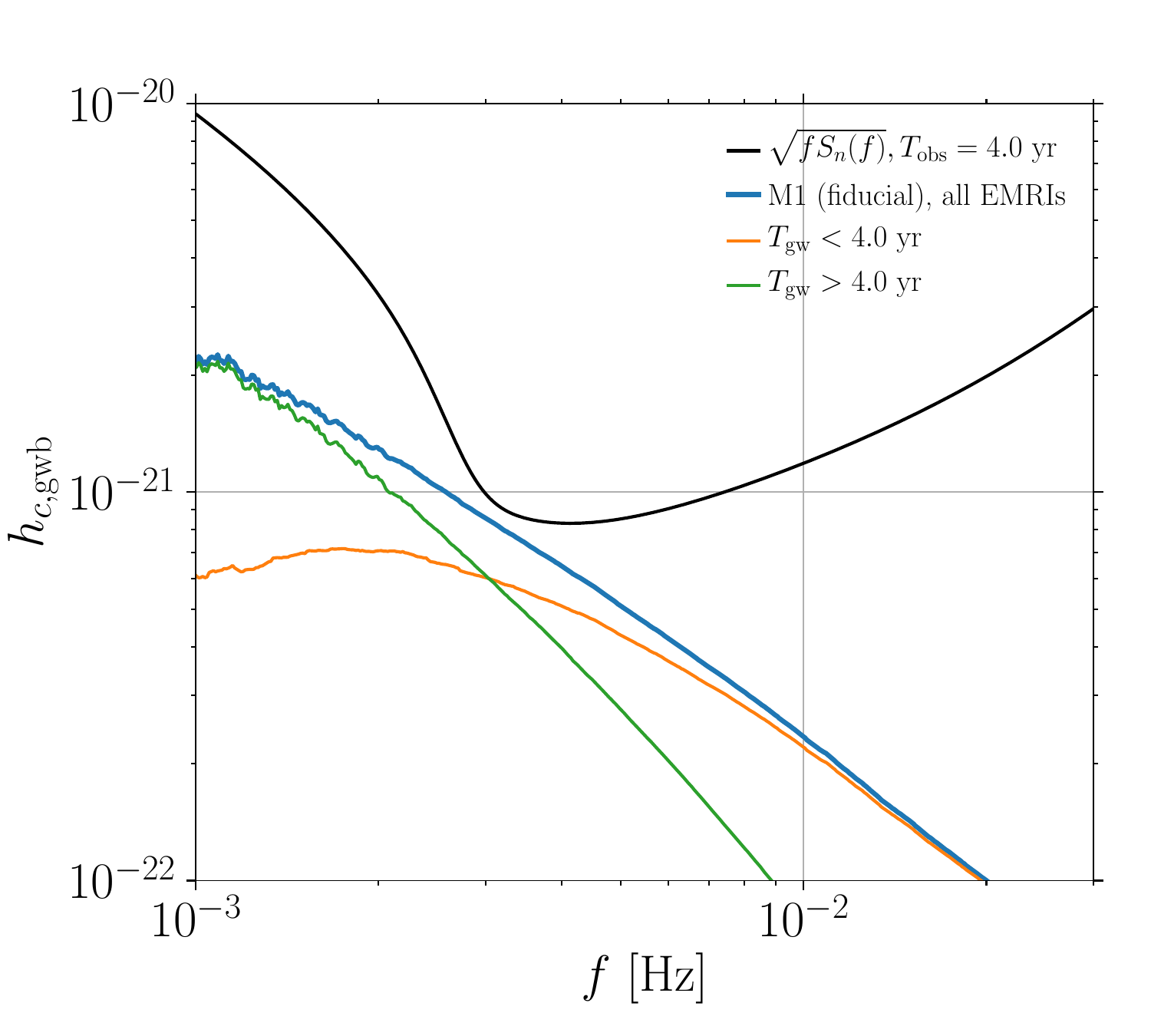}
    \caption{Characteristic strain of the GWB generated by EMRIs in the fiducial model (M1) that merge during the LISA mission (i.e. within four year, orange line) compared to that arising from non-coalescing sources (green line). The two sub-population build up to form the total GWB for M1 model (orange line, see also Fig.~\ref{fig:gwb_all1}).}
    \label{fig:gwb_inout}
\end{figure}

Focusing on the fiducial model M1, we now turn the discussion on the detailed contribution of individual EMRIs to the GWB build-up. In Fig.~\ref{fig:gwb_inout} we show how the characteristic strain of the GWB get shaped when considering EMRIs that do (orange line) and do not (green line) plunge during the $T_{\rm obs}$, here assumed to be four years. As expected, plunging sources dominate the high frequency end of the GWB. Below $3\,$mHz, however, the larger contribution to the signal comes from EMRIs that are still relatively far from final plunge. This confirms that considering the global population of EMRIs is important in order to asses the true level of the GWB, which would be otherwise underestimated if we account only for sources plunging during the mission lifetime. 
In fact, for this fiducial model, if we consider only EMRIs plunging within $T_{\rm obs}$, the resulting GWB would be well below the LISA instrumental noise, as shown by the orange curve. We would therefore erroneously predict that the EMRI GWB will not affect LISA sensitivity and the detectability of other sources.

\begin{figure}
    \centering
    \includegraphics[width=0.48\textwidth]{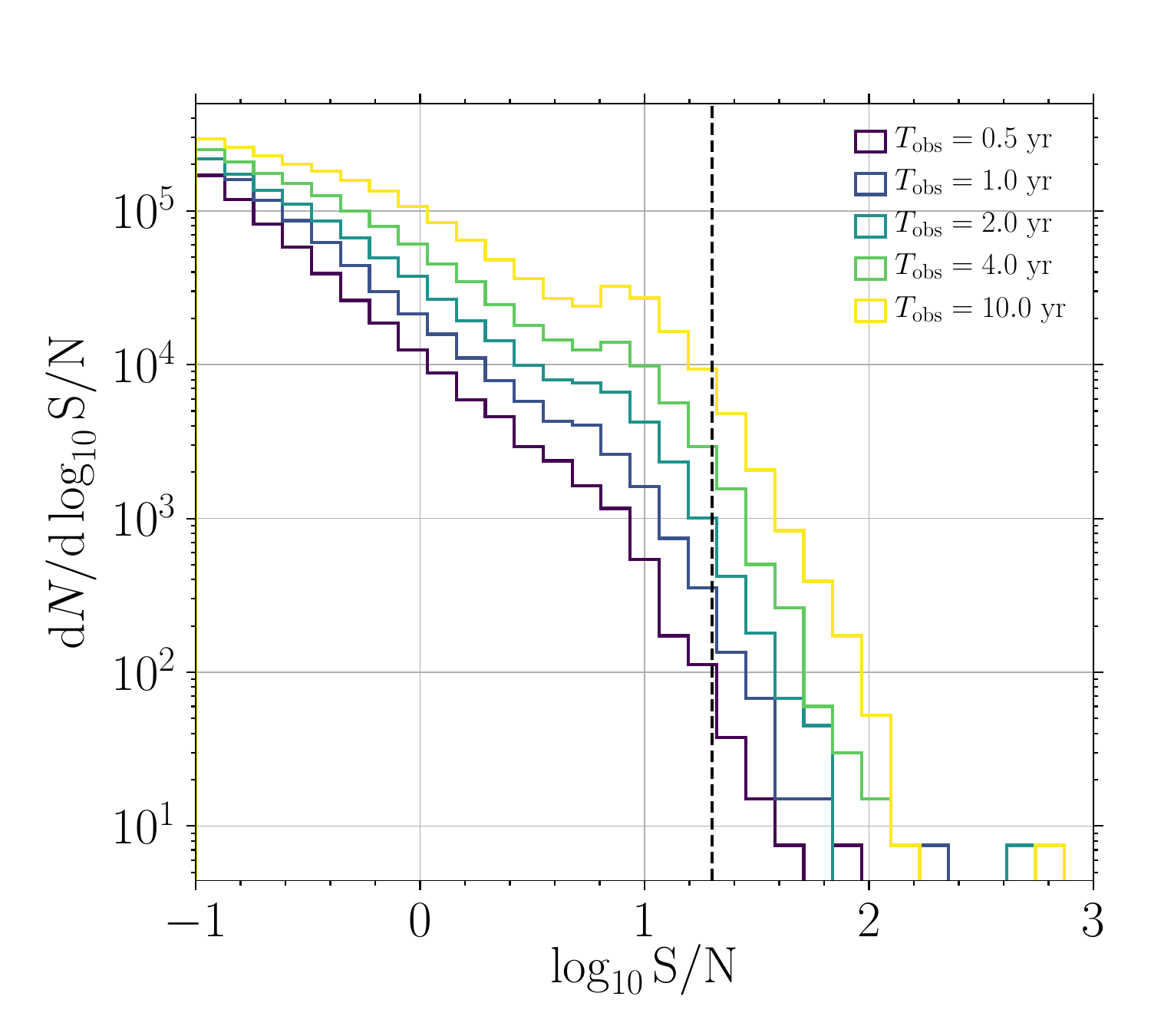}
    \caption{S/N distribution of EMRI for the fiducial model for different $T_{\rm obs}$ as labelled.}
    \label{fig:snr_distr}
\end{figure}

\begin{figure}
    \centering
    \includegraphics[width=0.48\textwidth]{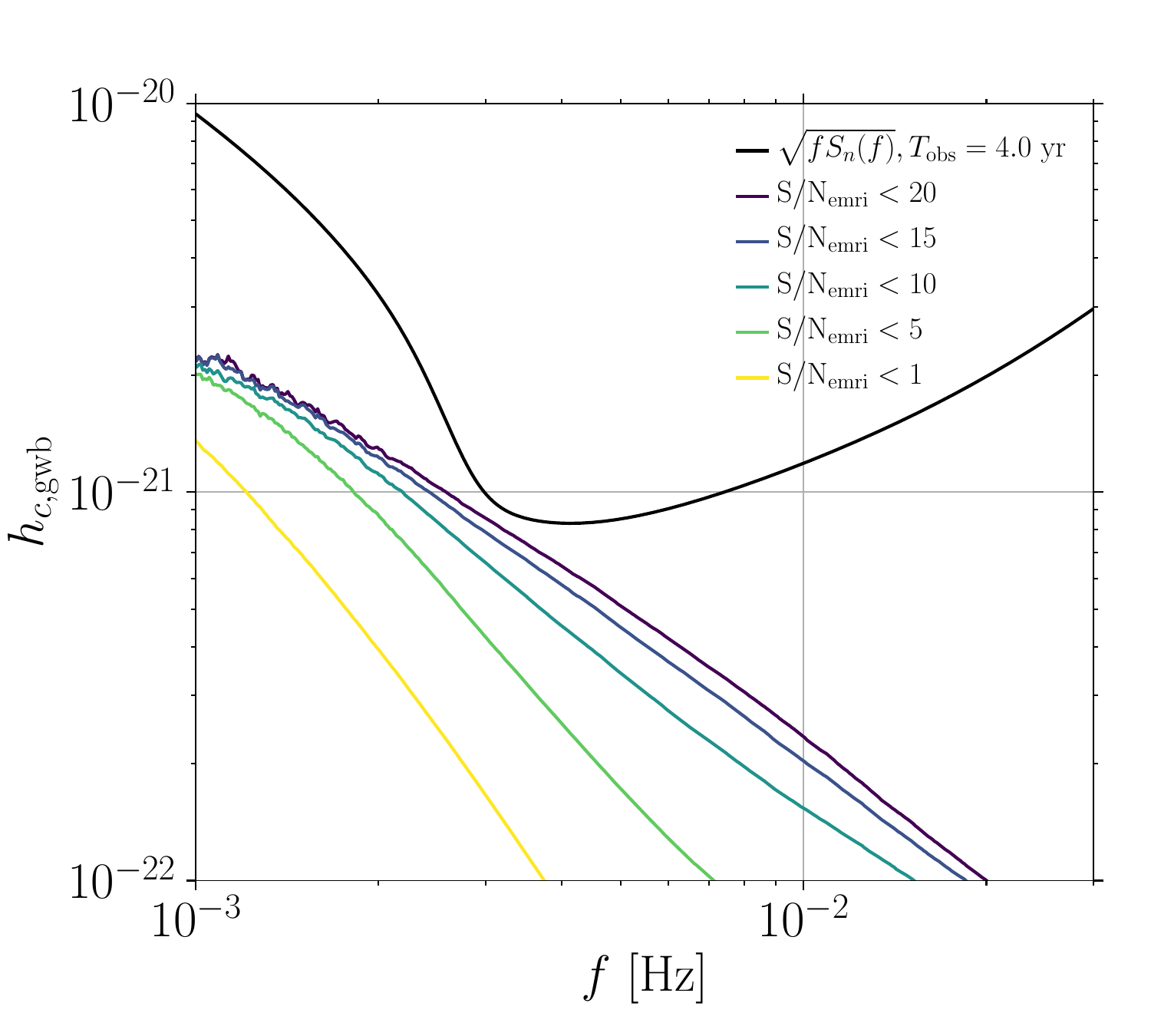}
    \caption{Characteristic strain of the GWB for the fiducial model (M1) when different S/N threshold for single source detection (and removal from the GWB) are considered.}
    \label{fig:snr_evol}
\end{figure}

So far, we considered a S/N threshold of 20 for individual EMRI detection (and subsequent signal removal). This number is backed both by theoretical considerations of waveform template counts and detectability threshold based on the resulting false alarm probability estimates, and by numerical simulations of signal injection and recovery. It should be noted that the mock LISA data challenge demonstrated the feasibility of extracting EMRIs down to ${\rm S/N}\approx 15$ in Gaussian instrumental noise \citep{2008CQGra..25r4026B,2010CQGra..27h4009B}. Although this is an highly idealized condition, it is nonetheless interesting to study the behavior of the GWB as a function of the threshold S/N considered for the sources contributing to it. In fact, since EMRI detection is inherently sensitivity limited, the distribution of the number of sources as a function of signal to noise ratio $\ud N/\ud({\rm S/N})$ is expected to have a steep dependence on S/N. For example, ignoring cosmological consideration, in the limit of Euclidean space and a population of similar sources uniformly distributed, since ${\rm S/N} \propto D^{-1}$ and $\ud N/\ud D\propto D^2$ (being $D$ the distance to the source), one expects $\ud N/\ud({\rm S/N})=(\ud N/\ud D)(\ud D/\ud({\rm S/N}))\propto{\rm S/N}^{-4}$. Although EMRIs are not all equal and are observed at cosmological distances, clearly this derivation does not strictly apply, still Fig.~\ref{fig:snr_distr} demonstrates that the above scaling is indeed an appropriate representation of the EMRI S/N distribution in our models. One can therefore expected the level of the GWB to be quite sensitive to the adopted S/N threshold for individual EMRI detectability. 
This is confirmed by Fig.~\ref{fig:snr_evol}, demonstrating that the characteristic strain of the background is dominated by high S/N sources. Sources with $10 < {\rm S/N} < 20$ contribute more then 50\% of the total GWB strain at $f=3\,$mHz. So if the detection S/N threshold is lowered to 10, than the level of the GWB would fall well below the LISA sensitivity curve, effectively eliminating (for this specific model) any residual confusion noise. Although this will be likely unfeasible, still the figure shows that the EMRI GWB is heavily dominated by slightly sub-threshold events, rather than stemming from the contribution of a vast number of dim sources.

\begin{figure*}
    \centering
    \includegraphics[width=0.48\textwidth]{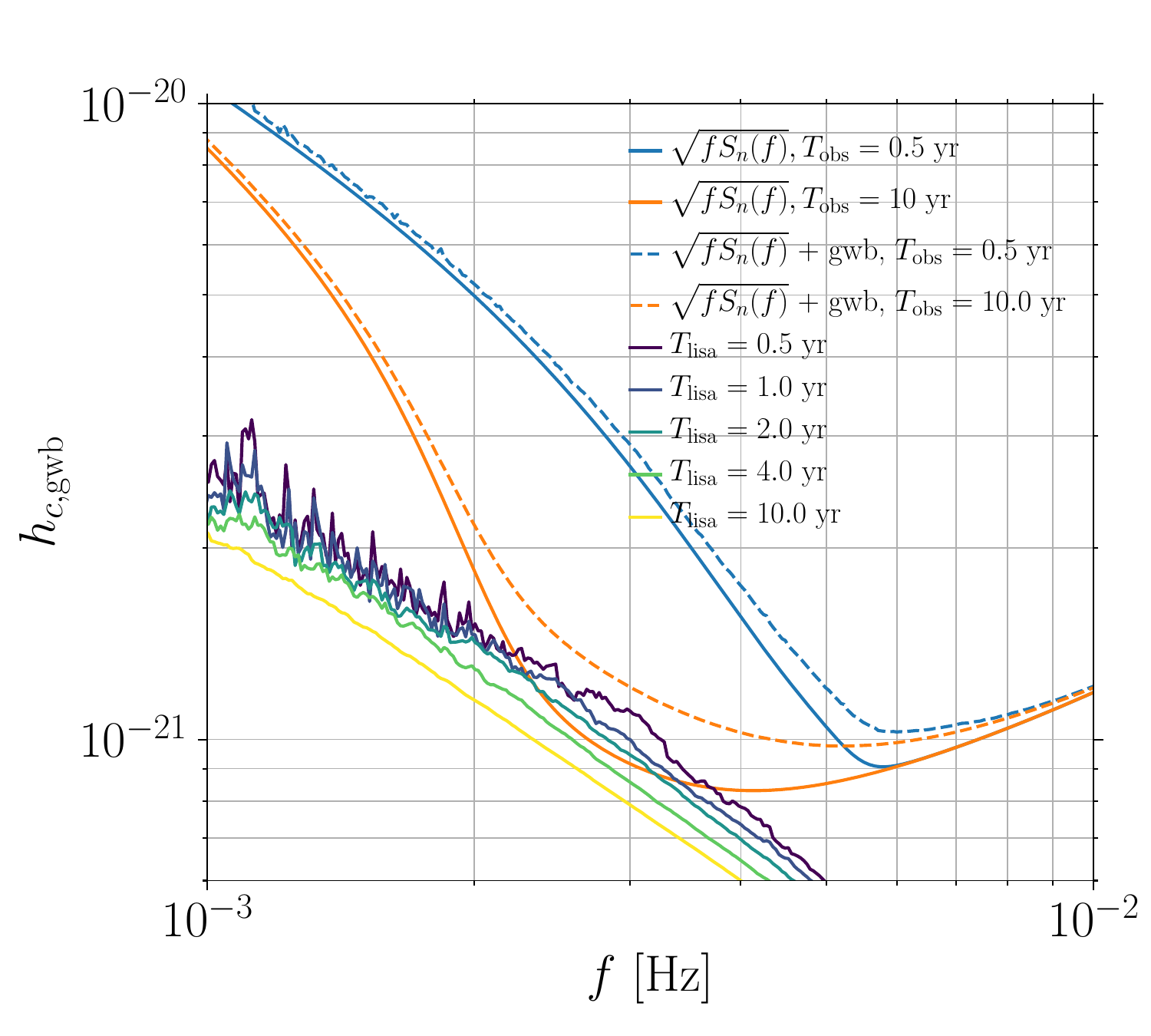}
    \includegraphics[width=0.48\textwidth]{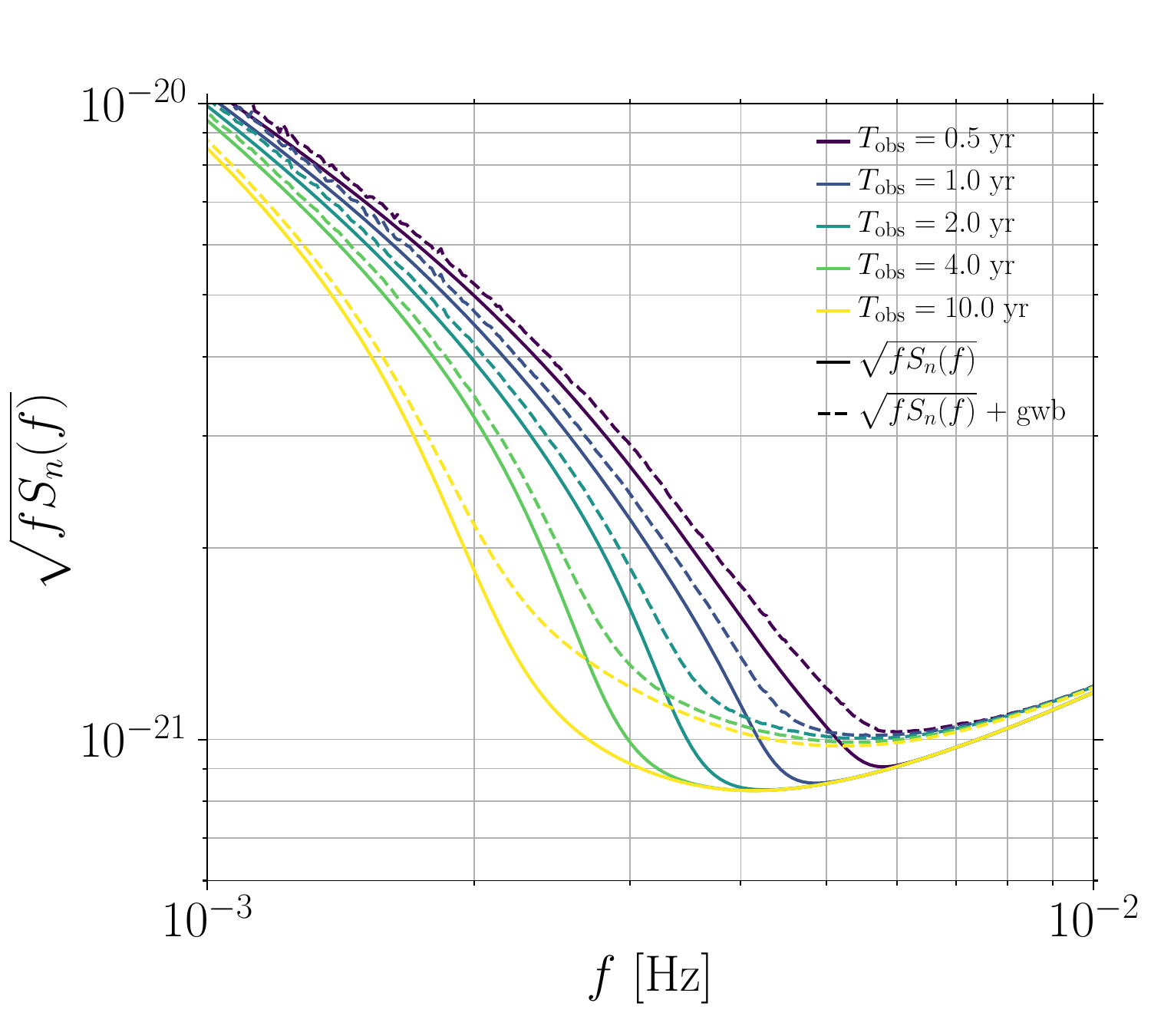}
    \caption{{\it Left panel}: Time evolution of the EMRI GWB for the fiducial model (M1) for different $T_{\rm obs}$ as labelled. As more and more sources exceed the S/N threshold those that can be resolved are removed from the GWB. Additionally, also the WD noise lowers with time (compare orange and blue curve) allowing more sources to be detected. {\it Right panel}: Time evolution of the LISA sensitivity curve accounting for WD noise (solid) and for the presence of the EMRI GWB (dashed).}
    \label{fig:gwb_evol}
\end{figure*}

Finally, the GWB level is expected to depend on $T_{\rm obs}$, mostly because EMRIs are long-lasting sources and longer observation times result in more sources building-up an S/N above the detectability threshold. This is depicted in the left panel of Fig.~\ref{fig:gwb_evol} which shows that the unresolved GWB in the bucket of the LISA sensitivity decreases by approximately 30\% as observations are accumulated from 6 months to 10 years. One would therefore naively expect that the residual EMRI GWB would have a stronger impact on the LISA sensitivity for shorter $T_{\rm obs}$. However, this is not the case, because when $T_{\rm obs}$ is shorter, the WD confusion noise is also more prominent in the mHz frequency range, dominating the LISA noise budget. Therefore, although the EMRI GWB is also higher, its relative impact on the LISA sensitivity is smaller compared to the WD confusion noise. This is shown in the right panel of Fig.~\ref{fig:gwb_evol}, where it is clear that the addition of the EMRI GWB has a stronger effect on the overall LISA sensitivity for longer $T_{\rm obs}$.\footnote{We remind the reader that we are ignoring the possibility of subtracting the WD confusion noise.} 
In terms of detectability, according to equation~\eqref{eq:power_SNR}, the GWB S/N should increase with $\sqrt{T_{\rm obs}}$. This, however, assumes $h_{c,{\rm gwb}}$ to be independent on the observation time. Since we found that $h_{c,{\rm gwb}}$ slightly decreases for longer $T_{\rm obs}$ due to the larger number of resolvable sources, the dependence of the GWB S/N is slightly flatter than  $\sqrt{T_{\rm obs}}$, as shown by the numbers in Tab.~\ref{tab:tab2}. For example, for M1 using the AKS waveform we find $\textrm{S/N}\propto T_{\rm obs}^{0.32}$.

\subsection{Properties of resolvable EMRIs}
\label{Sec:3.3}

\begin{figure}
    \centering
    \includegraphics[width=0.48\textwidth]{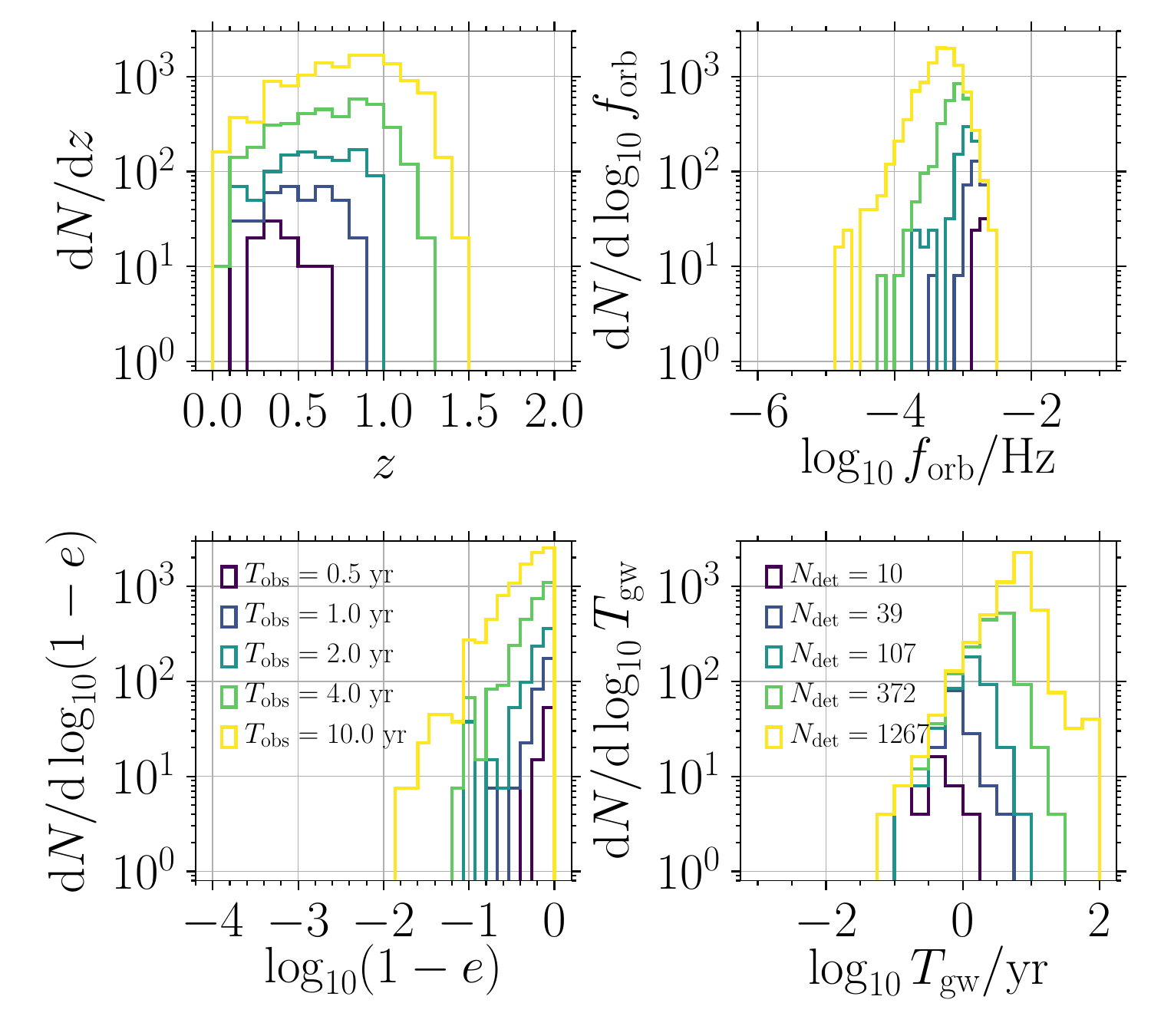}
    \caption{Differential number of detected EMRIs (${\rm S/N} > 20$) for the fiducial model (M1) as a function of redshift, initial orbital frequency, circularity and time to coalescence for different $T_{\rm obs}$ ranging from six months to ten years (see labels).}
    \label{fig:pop_time_evol}
\end{figure}

Finally, it is interesting to investigate the properties of individually resolvable EMRIs. Again, we take the fiducial M1 model, and consider the evolution of this source population as a function of $T_{\rm obs}$. In Fig.~\ref{fig:pop_time_evol} we report, from top left to bottom right, the distribution of detections as a function of redshift, initial orbital frequency and circularity (i.e. $1-e$) as well as the time to coalescence (indicated as $T_{\rm gw}$). In each plot, the various curves (from dark blue to yellow) refer to a different $T_{\rm obs}$ as labelled. In addition, in the bottom right panel, we also report the total number of detections (i.e. ${\rm S/N} > 20$) as a function of $T_{\rm obs}$. Here there are several things to notice. First, the number of resolvable EMRIs does not grow linearly with the observation time. To first order, persistent sources have ${\rm S/N}\propto T^{1/2}$ which results in a number of detected systems $N\propto T^{3/2}$ (assuming an Euclidean space and an homogeneous distribution of sources). EMRIs, however, sit at the intersection between persistent (e.g. WD binaries) and ``temporally localized'' sources (e.g. MBHB inspirals). Depending on their initial semi-major axis and eccentricity at the time of LISA observation, they can plunge in a matter of months or stay in band for the whole mission duration. Therefore, the dependence of the number of observed sources on the mission lifetime is not immediately obvious. Numbers reported in Fig.~\ref{fig:pop_time_evol} and Tab.~\ref{tab:tab2} suggest $N\propto T^{1.4}-T^{1.5}$, close to the expected scaling for persistent sources. 
This highlights the benefit of a longer mission duration, since the number of observed EMRIs grows faster then linearly with the observation time.

It is also interesting to see the evolution of the properties of observed sources as $T_{\rm obs}$ increases. Despite the redshift distribution of observed sources appears to evolve self-similarly with $T_{\rm obs}$ (top-left panel of Fig.~\ref{fig:pop_time_evol}), the same does not hold true for other properties. This is an obvious consequence of the shape of the LISA sensitivity curve combined with the S/N build-up of EMRIs. In fact, EMRIs tend to accumulate most S/N in the last months before plunge, when the signal is strong and tend to fall around the bucket of the LISA sensitivity curve (cf. Fig.~\ref{fig:hc_example}). Therefore, initially, for $T_{\rm obs}\leq 1\,$yr only systems with $f_{\rm orb}>10^{-3}\,$Hz tend to be resolvable (top-right panel of Fig.~\ref{fig:pop_time_evol}). Those are generally in a late evolutionary stage, and in fact their time-to-plunge when LISA starts observing is $<1\,$ yr and the sources plunge within the mission lifetime. As $T_{\rm obs}$ grows, systems with lower initial frequency and farther from final plunge have time to accumulate enough S/N to surpass the detection threshold, and consequently the $f_{\rm orb}$ ($T_{\rm gw}$) distribution extends to lower (longer) values. This also naturally results into a long tail of detected systems with high eccentricity. In fact, due to GW circularization, EMRIs detected within few months are emitting at high frequency and had already time to significantly circularize. When ``caught'' by LISA, their eccentricity is generally $<0.5$. However, as $T_{\rm obs}$ extends to 10 years, eccentric systems become increasingly important with as much as $5\%$ of all detection having $e>0.9$ and possibly few EMRIs with $e > 0.99$. Finally, it is interesting to notice that, despite there is a peak of observable EMRIs at $T_{\rm gw}\lesssim T_{\rm obs}$, there is also a long tail at $T_{\rm gw}>T_{\rm obs}$. In general, $5-10\%$ of EMRIs individually identified above the detectability threshold {\it will not} plunge within the LISA mission lifetime. For example, assuming the standard 4 year mission, there is an handful of EMRIs caught as early as $\approx 30\,$ yr prior to final plunge. This sub-population of ``non plunging EMRIs'' is generally overlooked when observable EMRI rates are computed in the literature (e.g. in Babak17). 

\section{Discussion}
\label{sec:discussion}

\subsection{Observational consequences of the EMRI GWB}

Our results have a number of theoretical as well as practical consequences. In fact, it is important to notice that in several variations of the standard scenario M1, the EMRI GWB will significantly contribute to the LISA noise budget around the sensitivity bucket, possibly jeopardising the detectability of other interesting sources. The effect might be particularly severe for two family of sources that are of paramount importance for the LISA science case, namely high redshift seed MBHBs and stellar origin BH (SOBH) binaries. 

LISA is expected to observe seed MBHBs as light as few $\times10^3\msun$ out to $z\approx 20$ \citep[see e.g.][]{2011PhRvD..83d4036S,2019MNRAS.486.4044B,Barausse2020}. These systems should naturally arise in cosmological scenarios where MBHs grow from low mass seeds, possibly remnants of population III (popIII) stars of few hundred solar masses \citep{2001ApJ...551L..27M}. During their early growth, the seeds are expected to form a large number of binaries, potentially detectable by LISA. Because of the low mass and high redshift, these sources slowly cross the bottom of the LISA sensitivity bucket and accumulate most of the S/N over several months \citep[cf. Fig.~2 in Ref.][]{2005ApJ...623...23S}. A degradation of about 50\% of the detector noise budget in this region will affect the mass and redshift threshold within which those systems are detectable, making more difficult to reconstruct the early MBH cosmic history. In the worst case scenario (which is optimistic from an EMRI detection stand point, M12), the sensitivity at $10^{-3}\,$Hz $<f<10^{-2}\,$Hz will be severely compromised, to the point that it might be impossible to distinguish among seed formation channels. 

Likewise, although to a lesser extent, the GWB will affect detectability of SOBH binaries \citep{2016PhRvL.116w1102S,2016MNRAS.462.2177K,2019PhRvD..99j3004G,2020PhRvD.102h4056M}. Multiband sources, crossing to the ground based detector band within few years, are expected to be observed at $f>0.01\,$Hz, and therefore should not be affected. There is, however, also a large population of slowly inspiralling systems falling at lower frequencies, which might be lost if a significant EMRI GWB contributes to the LISA noise budget at $f<0.01\,$Hz. This will also be a problem for some resolvable galactic WD binaries emitting at $f>1\,$mHz.

In light of the above considerations, from a theoretical standpoint it becomes of paramount importance to better understand EMRI rates. The three orders of magnitude uncertainties in the rate reported in Babak17 stem mostly from the poor knowledge of the low mass end of the MBH mass function and of the characteristic EMRI rates per individual MBH. The latter in particular are generally based on numerical simulations of Milky Way type galactic nuclei \citep{2011CQGra..28i4017A,2015ApJ...814...57M}, extrapolated at lower masses by appropriately scaling the nuclear stellar density and the associated relaxation time with the MBH mass. Moreover, they are affected by the uncertain estimates of the ratio of direct plunges to EMRIs occurrence. Our findings call for targeted dynamical modeling and relativistic numerical simulations of dense nuclei around MBHs in the $10^5-10^6\msun$ mass range, most relevant to LISA, to better pin down the complex dynamics of EMRIs and reduce the uncertainty in rate estimates.  

\subsection{Comparison with previous work}

To the best of our knowledge, so far the only detailed computation of EMRI background has been performed by BC04b. Their computation relied on a simplified piece-wise function describing the energy density emitted by an individual EMRI, coupled with a number of empirical estimates of the MBH mass function and a scaling relation for the EMRI rate per MBH (${\cal R} \propto M^{3/8}$). Moreover, the computation was done using the sensitivity curve for ``Classic LISA'' \citep{2000PhRvD..62f2001L}, relevant at the time. Besides updating the LISA curve to the current design, our calculations also rely on later developments in the study of EMRI dynamics resulting in different rates $\mathcal{R}\propto M^{-0.2}$ \citep{Hopman2005,2009CQGra..26i4034G}, a more detailed computation of the emission of each individual EMRI, and on realistic MBH mass functions. Most notably, this returns a flatter GWB spectrum; while at $10^{-3}\,$Hz $<f<10^{-2}\,$Hz BC04b finds a steep GWB with $h_c\propto f^{-1.5}$, our calculation results in $h_c\propto f^{-1}$. This is likely due to the larger contribution of EMRIs forming around lighter MBHs, contributing more power to the signal at high frequencies. Nonetheless, we confirm the main finding of BC04b that the EMRI GWB might significantly affect the LISA sensitivity in the bucket and should be taken into account seriously. 
\begin{figure}
    \centering
    \includegraphics[width=0.48\textwidth]{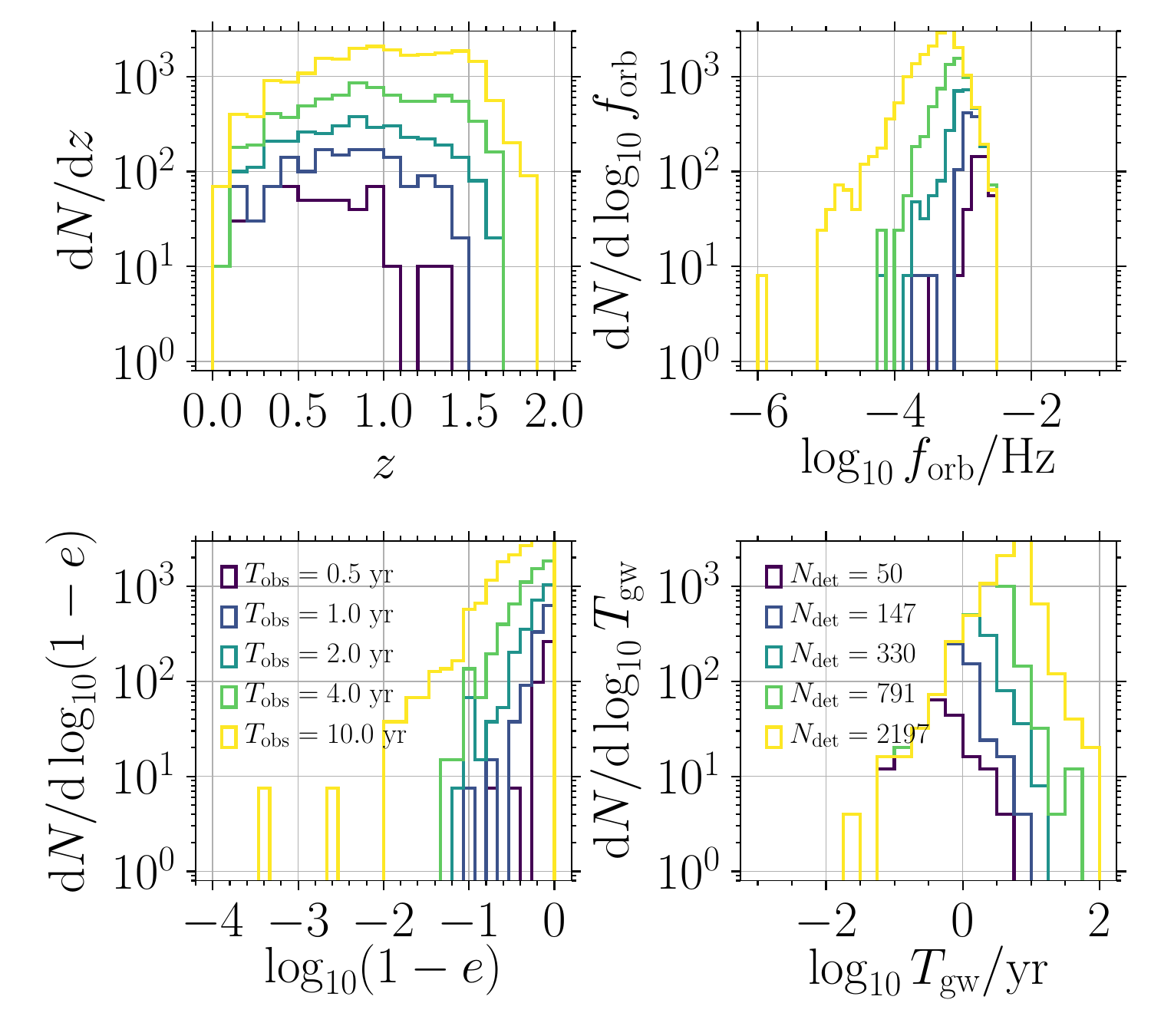}
    \caption{Same as Fig.~\ref{fig:pop_time_evol}, but considering considering the sensitivity curve of \citet{Babak2017}.}
    \label{fig:pop_time_evol_babak}
\end{figure}

We can also compare the number of detectable systems (${\rm S/N} > 20$) with those given in Babak17. Comparing numbers in Fig.~\ref{fig:pop_time_evol} and Tab.~\ref{tab:tab2} with those given in Tab.~\ref{tab:tab1}, our detection rates appear to be more than a factor of two smaller. The discrepancy can be ascribed both to the waveform model adopted, but also to the different sensitivity curve employed. To check the contribution of the latter, we repeated our calculations changing LISA sensitivity in order to match that used by Babak17. Results reported in Fig.~\ref{fig:pop_time_evol_babak} show that by doing this, we get rates that are larger by a factor of $\approx 2$. Detections per year reported in the figure are consistent with those given by the AKS waveform in Babak17 (last column of Tab.~\ref{tab:tab1}). We conclude that differences due to our simpler waveform modelling (most notably the use of inclination-polarization averaged fluxes, cf. equation~\ref{eq:hcn}) are minor. We also stress that a detailed comparison of the rates is not straightforward. In fact, Babak17 consider catalogs of EMRIs plunging over 10 years and select detectable ones by integrating {\it for each of them} the signal from the plunge backwards in time for two years. They then divide numbers by 10 to obtain a yearly detection rates. Although this procedure provides results that are ball-park correct, it does not correspond to any realistic observation scenario. Conversely, based on the same catalogs of plunging EMRIs, we consistently construct the differential $\ud^4 N_{\rm EMRI}/(\ud M \ud z \ud f \ud e)$ at the time LISA starts its observations and we integrate the signal of each event over $T_{\rm obs}$. This allows to: i) take into account for the lifetime of each EMRI in the simulation (e.g. the signals from EMRIs that are only few months from coalescence when LISA starts taking data are consistently integrated only for few months), and ii) identify EMRIs that can be resolved despite they do not plunge within $T_{\rm obs}$ (that can account for up to 10\% of the whole population, as described in Section~\ref{Sec:3.3}). The most important difference is that an ``universal EMRI rate'' per year cannot be technically defined regardless of the mission duration, since the number of detections does not scale linearly with $T_{\rm obs}$.

\subsection{Caveats related to the employed waveform}

Finally, it is worth pointing out some caveats, mostly related to the waveform used in this work. Our waveform model is a simplified version of the AK waveform constructed by BC04a, where instead of integrating the PN equations for the evolution of the orbital elements, we consider only the quadrupole fluxes given by \citet{Peters1964}. This was necessary to make use of the $f-e$ relation to speed-up the GWB computation, as described in Section~\ref{sec:SNR}. Nonetheless, the overall $h_c$ spectrum of individual EMRIs obtained in this way does not significantly different from a computation including all the PN orders considered by BC04a. Note, moreover, that by evaluating $h_c(f_n)$ at integer multiples of the Keplerian orbital frequency $f_n=n\times{f_{\rm orb}}$, we avoid the frequency mismatch of the original AK waveform of BC04a, as first noted in \citet{2015CQGra..32w2002C}.

\begin{figure}
    \centering
    \includegraphics[width=0.48\textwidth]{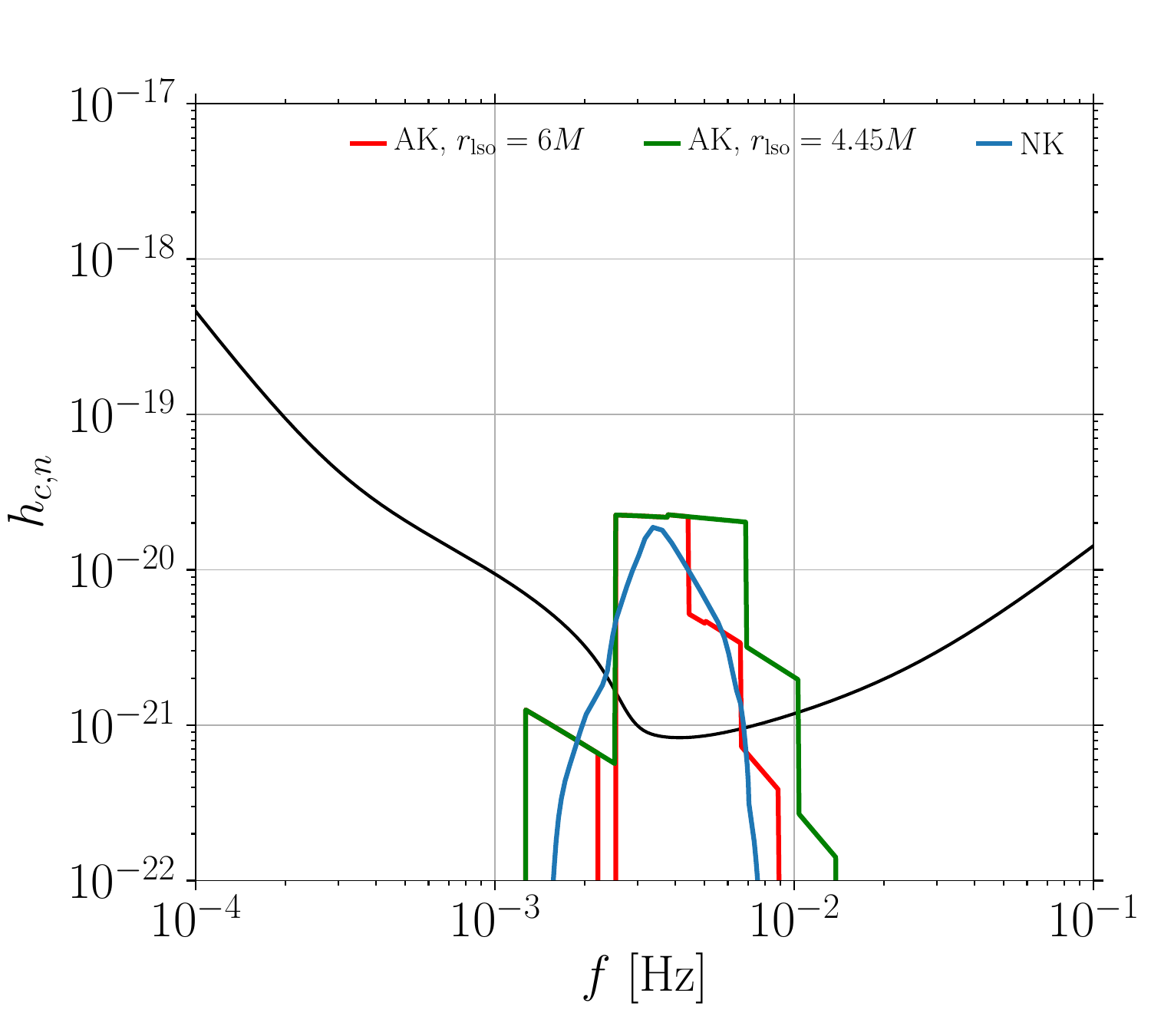}
    \caption{Revisiting of Fig.~3 of \citet{2017PhRvD..96d4005C}. The figure shows $h_c$ emitted by an EMRI with redshifted masses $10^6\msun, 10\msun$, primary spin parameter $a/M=0.5$ at $z=0.8$.
    The blue line shows the NK waveform, whereas the red and green lines show AK waveforms truncated at the Schwarzschild ISCO and at the Kerr ISCO for $a/M=0.5$ respectively.}
    \label{fig:cfr_NK}
\end{figure}

\citet{2017PhRvD..96d4005C} built on the work of BC04a to construct an ``augmented analytic kludge'' (AAK) waveform. AAK is a fast and efficient model able to match the waveform obtained by solving numerically the trajectory of the CO along the Kerr geodesic (also referred to as numerical kludge, NK). In Fig.~\ref{fig:cfr_NK}, we compare the NK waveform to our simplified AK model for an EMRI with redshifted masses of $10^6\msun+10\msun$, primary spin parameter $a/M=0.5$ at $z=0.8$. Our AK model captures the salient features of the waveform, but is not accurate in modelling the spectrum close to the final plunge, because it overestimates the frequency of the ISCO. In fact, for the considered $a/M=0.5$ system, the NK waveform (blue) frequency range is better matched by a Schwarzshild AK waveform (red) rather than an AK waveform with $a/M=0.5$ (green). We deem, however, this waveform inaccuracy unimportant for our purposes, since we found that the shape and amplitude of the GWB is essentially independent on the ISCO choice (cf. Section~\ref{sec:3.1}).
It should be also noted that the different shape and normalization of the waveforms stem from the fact that we are comparing the AK inclination-polarization averaged $h_c$ to a NK non-averaged $h_c$, where the inclination angle of the source is assumed to be $\iota=\pi/6$.

Finally, we stress that our GWB computation is based on the use of inclination-polarization averaged fluxes, and a more sophisticated model should take into account the inclination of each individual system with respect to the observer line of sight. We expect, however, this effect to produce only minor adjustments to our estimates.

\section{Conclusions}
\label{sec:conclusions}

We have performed a comprehensive study of the stochastic GWB produced by a cosmological population of EMRIs in the LISA band. Our work expands and updates the original investigation by BC04b. We built our computation on the state of the art plunging EMRI catalogs of Babak17, constructed from EMRI population models encompassing a range of physically motivated prescriptions for the most relevant ingredients affecting their formation, including: i) the cosmic evolution of the MBH mass function, ii) the relation between MBH mass and density of the surrounding stellar environment, iii) the modification of such environment due to galaxy mergers, iv) the rate of EMRI formation given the properties of the galactic nucleus and v) the occurrence ratio of direct plunges to EMRIs. We devised a formalism to construct from each catalog the distribution of EMRIs as a function of MBH mass, redshift, orbital frequency and eccentricity at the time LISA observations start, and we integrated their signals using a simplified, inclination-polarization averaged version of the AK waveform of BC04a. We removed signals with ${\rm S/N} > 20$ and computed the residual GWB by adding up the remaining sources. We then investigated in details the properties of resolvable sources and residual GWB as a function of the LISA mission duration $T_{\rm obs}$. Our main findings can be summarized as follows:  
\begin{itemize}
    \item there is about a factor 1.5 dex uncertainty in the expected GWB level, consistent with the 3 dex uncertainty in the EMRI rate estimated by Babak17 (cf. Fig.~\ref{fig:gwb_all1});
    \item the EMRI GWB is easily detectable for most considered astrophysical scenarios, with S/N of several hundreds to several thousands. Even in the most pessimistic case marginal detection is expected with S/N of a few (cf. Tab.~\ref{tab:tab2});
    \item most of the investigated models produce a residual GWB that is going to affect the LISA sensitivity curve in the bucket. In particular, the GWB level predicted by our fiducial model M1 grazes the LISA noise curve around $3\,$mHz, whereas in the optimistic (in terms of EMRI counts) model M12, the residual GWB is more than a factor of 3 higher then the instrument noise at the same frequency, effectively erasing the LISA sensitivity bucket (cf. Fig.~\ref{fig:gwb_all2});
    \item those results are largely insensitive to the spin distribution of MBHs and to the details of the adopted waveform close to final plunge (cf. Fig.~\ref{fig:gwb_spin_nospin});
    \item the GWB mildly decreases for longer mission durations, dropping by about 20-25\% as the observation time increases from 6 months to 10 years (cf. Fig.~\ref{fig:pop_time_evol});
    \item the dominant contribution to the GWB is produced by EMRIs with $10 < {\rm S/N} < 20$, just below the ${\rm S/N} = 20$ detectability threshold (cf. Fig.~\ref{fig:snr_evol}) and at $f<3\,$mHz is due to EMRIs that {\it will not} plunge within the mission lifetime (cf. Fig.~\ref{fig:gwb_inout});
    \item the number of individual EMRIs observable with ${\rm S/N} > 20$ grows as $T_{\rm obs}^{1.4}-T_{\rm obs}^{1/5}$, i.e. faster than linearly;
    \item as many as 10\% of resolved EMRIs do not plunge within the LISA observation time (cf. bottom-right panel of Fig.~\ref{fig:pop_time_evol}).
\end{itemize}

We compared our findings to BC04b and Babak17, discussing the most relevant differences, and we highlighted their implications for LISA. In particular, if the actual EMRI rates are on the high side of the estimated range, the sensitivity of LISA between $10^{-3}\,$Hz and $10^{-2}\,$Hz can be severely compromised, with undesired consequences particularly for the detection of high redshift, low mass seed MBHBs and of SOBH binaries, which are both primary target sources of LISA. This calls for a concerted theoretical effort aimed at a better understanding of EMRI formation and dynamics, with the goal of reducing the cosmic EMRI rate uncertainty range.  

Finally, we discussed a number of caveats mostly related to the use of a simplified AK waveform, noticing however that our main results should be robust against this assumption. Our GWB computation assumes that all EMRIs with ${\rm S/N} > 20$ can be correctly identified and accurately subtracted from the LISA data stream. Eventually, this requires a faithful model of the EMRI waveform, which will likely require the full development of second order self force computations, currently under way \citep[e.g.][]{2009CQGra..26u3001B,2012PhRvD..85f1501W,2020PhRvL.124b1101P}. Any mismatch between the true signal and the waveform model used in the analysis will result in imperfect removal of resolved signals, spuriously increasing the residual unresolved GWB. Moreover, our analysis assume no subtraction of the Galactic WD confusion noise: a consistent detection pipeline for an EMRI GWB will have to consistently model the time evolution of the WD confusion noise, including the possibility of its (at least partial) subtraction due to its anisotropic nature.
It is therefore of paramount important for the full scientific success of the LISA mission to pursue a concerted community effort aimed at a better understanding of all aspects of EMRIs, from the intricacies of their astrophysical and dynamical origin responsible for their expected cosmic rate, to the finest details of their emitted gravitational waveforms as well as the complexity of building end-to-end detection pipelines.


\begin{acknowledgments}
    We thank Stanislav Babak, Nikolaos Karnesis, Alvin Chua and Simeon Bird for valuable discussions and suggestions.
    Numerical calculations have been made possible through a CINECA-INFN agreement, providing access to resources on GALILEO and MARCONI at CINECA. A.S. is supported by the European Research Council (ERC) under the European Union’s Horizon 2020 research and innovation program ERC-2018-COG under grant agreement No. 818691 (B Massive).
\end{acknowledgments}





\bibliography{biblio}

\end{document}